\PassOptionsToPackage{table}{xcolor}

\documentclass[AMA,STIX1COL]{WileyNJD-v2}

\articletype{}

\received{26 April 2016}
\revised{10 June 2016}
\accepted{10 June 2016}

\usepackage{fix-cm}
\DeclareSymbolFont{CMletters}{OML}{cmm}{m}{it}
\DeclareMathSymbol{\ell}{\mathord}{CMletters}{"60}

\newcommand{\E}{\thinspace}                                
\newcommand{\hs}[1]{\hspace{#1}}                           
\newcommand{\red}[1]{ {\scalebox{0.59} {$#1$}}}            
\newcommand{\lar}[1]{ {\scalebox{0.70} {$#1$}}}            
\newcommand{\bds}[1]{{\boldsymbol{#1}}}                    
\newcommand{\tb}[1]{{\textbf{#1}}}                         
\newcommand{\tx}[1]{{\text{#1}}}                           
\newcommand{\cx}[1]{{\texttt{#1}}}                         
\newcommand{\CT}{\multicolumn{1}{c}{\text{--}}}            
\newcommand{\B}{\beta}                                     
\newcommand{\bz}{\tb{z}}                                   
\newcommand{\lm}{\lar{\ell m}}                             
\newcommand{\G}{\gamma}                                    
\newcommand{\tm}{\times}                                   
\newcommand{\pd}{{\scalebox{0.80}{\E$\times$\E}}}          
\newcommand{\tsp}{$t_\red{\raisebox{0.4ex}{\tx{SP}}}$}     

\newcommand{\rulefiller}{
  \arrayrulecolor{gray!40}
  \specialrule{0.275ex}{0pt}{-0.275ex} 
  \arrayrulecolor{black}
}

\makeatletter
\newcommand{\raisemath}[1]{\mathpalette{\raisem@th{#1}}}
\newcommand{\raisem@th}[3]{\raisebox{#1}{$#2#3$}}
\makeatother

\usepackage{multirow, booktabs}
\usepackage[english]{babel}

\addto{\captionUSsenglish}{}
\usepackage[ansinew]{inputenc}

\begin{document}

\title{Semi-Markov multistate modeling approaches for multicohort 
event history data}

\author[1]{Xavier Piulachs*}
\author[2]{Klaus Langohr}
\author[3]{Mireia Besal\'u}
\author[4,5]{Natalia Pallar\`es}
\author[4,5,6]{Jordi Carratal\`a}
\author[4,5]{Cristian Teb\'e}
\author[2]{Guadalupe G\'omez Melis}

\authormark{ }
  
\address[1]{\orgdiv{Department of Statistics and Operations Research}, \orgname{Polytechnic University of Catalonia}, 
  \orgaddress{\state{Campus Terrassa}, \country{Spain}}}
 \address[2]{\orgdiv{Department of Statistics and Operations Research}, \orgname{Polytechnic University of Catalonia}, 
  \orgaddress{\state{Campus Barcelona}, \country{Spain}}}
 \address[3]{\orgdiv{Department of Genetics, Microbiology and Statistics}, \orgname{University of Barcelona}, 
  \orgaddress{\state{Barcelona}, \country{Spain}}}
\address[4]{\orgdiv{Bellvitge Biomedical Research Institute}, \orgname{Bellvitge University Hospital}, 
  \orgaddress{\state{L'Hospitalet de Llobregat}, \country{Spain}}}
\address[5]{\orgdiv{Department of Clinical Sciences}, \orgname{University of Barcelona}, 
  \orgaddress{\state{L'Hospitalet de Llobregat}, \country{Spain}}}
\address[6]{\orgdiv{Department of Infectious Diseases}, \orgname{Bellvitge University Hospital}, 
  \orgaddress{\state{L'Hospitalet de Llobregat}, \country{Spain}}}

\corres{*Xavier Piulachs, Department of Statistics and Operations Research, Polytechnic University of Catalonia, Campus Terrassa, 
       Building TR5, C. Colom 11, 08222 Terrassa, Spain. \email{xavier.piulachs@upc.edu}}

\abstract[Summary]{
Two Cox-based multistate modeling approaches are compared 
for modeling a complex multicohort event history process. 
The first approach incorporates cohort information as a 
fixed covariate, thereby providing a direct estimation of 
the cohort-specific effects. The second approach includes 
the cohort as stratum variable, thus providing an extra 
flexibility in estimating the transition probabilities. 
Additionally, both approaches may include possible 
interaction terms between the cohort and a given prognostic 
predictor. Furthermore, the Markov property conditional 
on observed prognostic covariates is assessed using a 
global score test. Whenever departures from the Markovian 
assumption are revealed for a given transition, the time 
of entry into the current state is incorporated as a 
fixed covariate, yielding a semi-Markov process. The two 
proposed methods are applied to a three-wave dataset of 
\mbox{COVID-19}-hospitalized adults in the southern Barcelona 
metropolitan area (Spain), and the corresponding performance 
is discussed. While both semi-Markovian approaches are shown 
to be useful, the preferred one will depend on the focus of 
the inference. To summarize, the cohort-covariate approach 
enables an insightful discussion on the the behavior of the 
cohort effects, whereas the stratum-cohort approach provides 
flexibility to estimate transition-specific underlying risks 
according with the different cohorts.}


\maketitle

\section{Introduction} \label{sec1}
Within the scope of biomedical studies, there is a rising 
interest in assessing the elapsed time from some predefined 
origin until the occurrence of a given terminal event among 
several mutually exclusive ones, competing with each other. 
As a natural extension of this competing risks setting, the 
analysis of intermediate clinical states over the follow-up 
period can also be addressed, thus giving a more 
comprehensive picture of a subject's trajectory through 
distinct states of a particular real-world complex process. 
This leads to modeling consecutively recorded event times, 
for which multistate models provide a natural framework 
(Andersen \& Keiding, 2002; Houwelingen \& Putter, 2008; 
Meira-Machado et al., 2009; Geskus 2015; Cook \& Lawless, 
2018).~These models, typically specified in terms of the 
transition-specific hazards between subsequent states, 
allow us to enhance predictive performance regarding the 
probability of being in a particular state at a certain time. 

This article describes two Cox-based multistate modeling 
approaches, termed M1 and M2, to analyze event history data 
that involves two commonly encountered technical difficulties. 
On the one hand, the typically adopted Markovian assumption 
for a given transition is not necessarily valid and needs 
to be checked. To this end, a global testing procedure for 
the Markov property is applied, so historical effects may 
be included in both the M1 and M2 approaches whenever a 
non-Markovian context is proven. On the other hand, the 
behavior of many infectious diseases may depend on temporal 
and population patterns. For instance, seasonal influenza 
may progressively affect different population cohorts over 
time (Darroch \& McCloud, 1990),~while Alzheimer's disease 
affects distinct population groups differently (Birkenbihl 
et al., 2022).~Moreover, disease outcomes may even vary 
based on the medical center at which the condition is 
treated (Freijser et al., 2023).~In order for the unobserved 
cohort features influencing a given transition hazard to be 
properly incorporated, each of the two approaches introduces 
an alternative strategy. The M1~approach advocates including 
a cohort covariate, this becoming a particularly 
useful alternative when considering a reduced number of 
cohorts (Gelman \& Hill, 2007).~However, the cohort effect 
itself sometimes cannot be assigned a direct interpretation, 
with researchers being more interested in explaining how 
the impact of a baseline prognostic covariate changes 
with each cohort. In this case, an additional interaction 
term between the cohort and the covariate of interest 
should be added. Alternatively, assuming that the effect 
of observed prognostic covariates is of the highest interest, 
the M2~approach consists of stratifying a given transition 
hazard according to the disjoint cohorts. Here, a separate 
transition-specific baseline hazard function is estimated 
for each cohort, while preserving common coefficient 
estimates across cohorts for the prognostic covariates. 
As the price paid for stratification, the cohort effect 
is no longer included. On the other hand, estimates for 
the remaining covariates are more robust than in the 
M1~approach, though at the expense of some loss of 
efficiency (particularly when estimating transition 
probabilities between two states). Moreover, the 
interaction between the cohort and a particular 
covariate can still be assessed.

To illustrate the suggested modeling approaches, medical 
data on subjects hospitalized for severe coronavirus 
disease (\mbox{COVID-19}) is analyzed. Particularly, our 
applied research comes from the project entitled {\it 
Dynamic evaluation of~\mbox{COVID-19}~clinical states 
and their prognostic factors to improve intra-hospital 
patient management} (Divine project), this data complying 
with the principles of the Declaration of Helsinki. 
The~\mbox{COVID-19}~process comprises a finite number of 
intermediate, mutually-exclusive disease severity states, 
ranging from asymptomatic to critically ill subjects who 
require hospitalization (Huang et al., 2020; Chan et al., 
2020).~Among the latter, the course of \mbox{COVID-19} 
can induce severe pneumonia, which can lead to the need 
for ventilatory support or even to death. Additionally, 
this pathological process involved a significant 
heterogeneity in the degree of severity between subjects, 
as the waves of the pandemic ebbed and flowed over time 
(Carbonell et al., 2021; Buttenschon et al., 2022). As 
part of the Divine project, it is of major interest to 
elucidate subject-specific changes in mortality trends 
when analyzing data from the first three \mbox{COVID-19} 
waves in Spain. To avoid selection bias, only subjects 
who were infected before hospital entry were selected, 
as well as subjects without a predetermined ceiling of 
care (i.e., those deemed suitable to receive advanced 
care). Further, virtually none of the subjects under 
study had received a \mbox{COVID-19} vaccine prior to 
hospital admission. Although a considerable proportion 
of the world's population is now vaccinated, an added 
value of our observational data is that it provides a 
comprehensive multistate modeling framework to understand 
and compare the information from the three distinct 
\mbox{COVID-19} waves that occurred during the most 
complicated time period of the disease. 

Our two Cox-based multistate modeling approaches are 
proposed to (i) assess the Markov hypothesis for all 
the relevant transitions in a multistate process, and 
(ii) quantify how cohort membership interacts with some 
baseline prognostic covariate when modeling a given 
transition hazard to the next state. Basically, both 
approaches account for past-history influence whenever 
the Markovian condition does not hold, while respectively 
applying the above mentioned strategies to incorporate 
the influence of the cohort. Each strategy has its 
strengths and weaknesses, and determines the type of 
information that may be derived from the approach. The 
remainder of the paper proceeds as follows. Section~2 
introduces the motivating study of the methods presented, 
consisting of a large longitudinal dataset of adults with 
confirmed \mbox{COVID-19} infections who were hospitalized 
during one of the first three coronavirus waves in the 
southern Barcelona metropolitan area. Section~3 describes 
the M1 and M2 approaches for modeling a given transition 
hazard. Section~4 illustrates the application of our two 
approaches to the motivating event history data. Lastly, 
Section~5 discusses the main results and suggests some 
possible directions for future research.
\section{Three-wave COVID-19 data in the southern Barcelona urban area}
The motivating three-wave data, which will be referred 
to as the \textsf{div3W}~data, covers a part of the 
database from the Divine project and comprises~$n=$~3290~
\mbox{COVID-19-hospitalized}~adults without a ceiling of 
care. They were admitted during the first three \mbox{COVID-19}  
waves occurring in Spain, spanning from March 2020 to February 
2021, to five collaborating medical centers located in the 
southern Barcelona metropolitan area. There were 2074 subjects 
recruited during the first wave, March--April 2020, 611 
subjects during the second wave, October--November 2020, 
and 605 subjects during the third wave, January--February 
2021. None of the individuals were lost to follow-up after 
hospital admission, so they were all monitored until either 
hospital discharge or death. Figure~1 shows the unidirectional 
multistate scheme used for modeling subject flow in the 
\mbox{COVID-19} hospitalization process across the three waves, 
together with the number of individuals in each wave passing 
from one state to other. Upon being admitted to hospital 
(state 0), subjects are immediately assigned to one of the two 
possible initial states according to the severity of disease: 
no severe pneumonia, NSP (state 1) and severe pneumonia, SP 
(state 2); thus, all hospitalized subjects are in one of these 
two states at $t = 0$. Subjects can then pass through three 
intermediate disease states: recovery (state 3), noninvasive 
mechanical ventilation, NIMV (state 4), and invasive mechanical 
ventilation, IMV (state 5). Finally, subjects can move to only 
one of two final states: discharge (state 6) or death (state 7). 
The following are particularly important features of this 
scheme. First, our data does not report the exit times from 
the SP, NIMV and IMV states to the recovery state. We therefore 
do not know the exact time point at which the subjects returned 
to the hospital ward before being subsequently discharged. To 
remedy this, these individuals are assigned a fixed 2-day stay 
in recovery, this period being the minimum transition time 
required from a medical standpoint. Second, a very small number 
of subjects are reported in at least one of the waves for the 
SP~$\to$~death, NSP~$\to$~death, recovery~$\to$~death, and 
NIMV~$\to$~death transitions; indeed, the small quantities 
observed in the first wave are even lower in the subsequent 
ones. These transitions correspond to individuals who died
prematurely either in the early and noncritical states, or 
without undergoing IMV, so in a way they experienced an 
unnatural in-hospital trajectory. Even though the number of 
these subjects is not particularly large across the different 
waves, it became lower as the pandemic unfolded and disease 
knowledge increased. Third, the subject's entry into the 
intensive care unit is not considered here as a state itself. 
The reason is that each medical center established its own 
criterion for ICU assignment according to the bed occupancy 
level, so a hypothetical ICU state would involve as many 
criteria as hospitals considered, possibly yielding misleading 
conclusions.
\begin{figure}[!ht]
\begin{center}
\includegraphics[scale = 0.275]{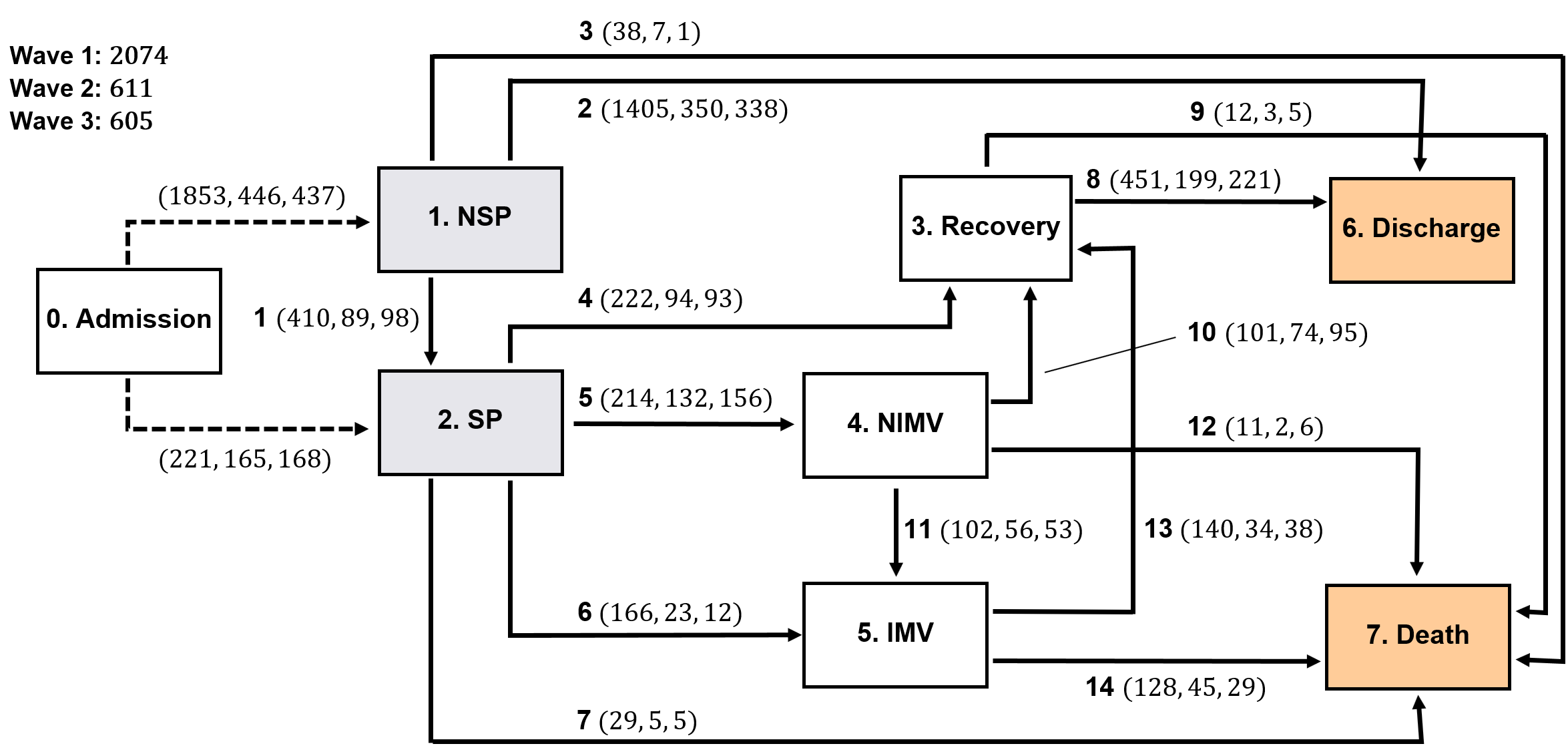}
\vspace{-0.10cm}
\end{center}
\caption{Multistate scheme adopted for modeling the clinical 
trajectory of the \mbox{COVID-19-hospitalized} subjects from 
the \mbox{\textsf{div3W}}~data. The in-hospital course 
of~\mbox{COVID-19} disease is divided for each wave into seven 
states after hospital admission: 1) no severe pneumonia (NSP), 
2) severe pneumonia (SP), 3) recovery, 4) noninvasive mechanical 
ventilation (NIMV), 5) invasive mechanical ventilation (IMV), 
6) discharge, and 7) death. The model envisages 14 possible 
transitions (solid arrows) between a pair of states, while it 
implicitly accounts for two instantaneous transitions (dashed 
arrows) from admission to either NSP or SP. Within each wave, 
the number of subjects observed per transition is indicated. 
\label{F1}} 
\vspace{-0.05cm}
\end{figure}
\par The \textsf{div3W}~data also provides a number of demographic 
and medical characteristics recorded upon hospital admission. Table~1 
summarizes the characteristics of the baseline prognostic covariates 
selected in our study following expert consultation, all assumed to 
be independently associated with the in-hospital mortality risk 
during each wave. We refer to Pallarès et al.~(2023)~for further 
details regarding these covariates, which are in line with the 
most frequently used information in the \mbox{COVID-19} literature 
(Wendel-Garcia et al. 2021, Berenguer et al. 2021).~Together with 
sex and age variables, the value of the safi ratio (Catoire et al., 
2021)~is used as illness severity scores. The safi value describes the 
relationship between the peripheral arterial oxygen saturation, 
typically ranging from 95\% to 100\%, and the fractional inspired 
oxygen concentration, with a value of 0.21 at room air. A better 
health condition is then related to higher safi values, with the 
interval [300, 476.2] covering most of the subjects considered. 

Across the three waves, more than half of the subjects are men, and 
a slight increase in age is observed as the pandemic progressed. It 
is also found that the proportion of subjects affected by SP is 
drastically higher in the later waves, and there is also a sharp 
increase in the proportion of subjects moving from SP to either NIMV 
or IMV; the latter account for 380 subjects in the first wave (18.3\%), 
155 subjects in the second wave (25.4\%), and 168 subjects in the 
third wave (27.8\%). These observed trends in age and disease severity, 
that might seem to behave in a counterintuitive way, can be explained 
by a relaxation in the ceiling-of-care criteria as the pandemic evolved, 
leading to a progressive increase in more critically ill subjects being  
admitted to hospital. Thus, as the strong restrictions during the first 
wave gradually disappeared, the subsequent waves saw a  higher admission 
rate for elderly subjects to the healthcare system; these subjects are 
the most seriously affected by the disease and often have multiple 
comorbidities. By contrast, a larger proportion of subjects died in the 
hospital during the first and second waves than during the third one.
\begin{table}[ht]
\centering\setlength{\aboverulesep}{0pt}\setlength{\belowrulesep}{0pt}
\setlength{\extrarowheight}{1pt}
\caption{Baseline characteristics of \mbox{COVID-19-hospitalized} 
  subjects from the \textsf{div3W}~data throughout the first, 
  second, and third pandemic waves. Categorical covariates are 
  expressed as counts and percentages (\%), whereas continuous 
  covariates are expressed as medians and interquartile ranges 
  (IQR). The selected baseline covariates are sex (female 
  indicator), age (years), and the safi value (units).\label{T1}}
\begin{tabular}{lccc}
\rowcolor{gray!40}
\tb{Characteristics}      & \tb{Wave~$\bds{1~(n=2074)}$} & \tb{Wave~$\bds{2~(n=611)}$} & \tb{Wave~$\bds{3~(n=605)}$}  \\
\midrule[\heavyrulewidth]
Baseline covariates                                                                                  \\
Sex (female)              &      854 (41.2\%)      &     222 (36.3\%)       &    248 (40.1\%)        \\
Age (years)               &     59 (49 -- 69)      &    62 (53 -- 71)       &     63 (52 -- 72)      \\
Safi value (units)        & 447.6 (395.2 -- 461.9) & 442.9 (346.4 -- 457.1) & 438.1 (342.9 -- 457.1) \\[1pt]
\rowcolor{gray!20}
Illness time course       &                        &                        &                        \\
\rowcolor{gray!20}
SP                        &        631 (30.4\%)    &     254 (41.6\%)       &    264 (43.8\%)        \\
\rowcolor{gray!20}
NIMV                      &        214 (10.3\%)    &     132 (21.6\%)       &    156 (25.8\%)        \\
\rowcolor{gray!20}
IMV                       &        268 (12.9\%)    &      79 (12.9\%)       &     65 (10.7\%)         \\
\rowcolor{gray!20}
In-hospital mortality     &        218 (10.5\%)    &      62 (10.1\%)       &     46 (7.6\%)       
\end{tabular} \\[-7.5pt]
\end{table}
\section{Semi-Markov multistate modeling for multiple cohorts}
\subsection{General framework}
Consider a finite set of mutually-exclusive states $\mathcal{R} = 
\{1, \dotsc, R \}$,  $R \in \mathbb{N}$, and let $\{X_i(t), \E t 
\geq 0 \}$, $i = 1, \dotsc, n$, denote the stochastic process that 
indicates the state occupied by the $i$th subject at a particular 
time point $t$ within a bounded time interval of interest $[0, \tau]$. 
The random process $X_i(t)$ is allowed to have independent, 
right-continuous sample trajectories on $[0, \infty)$. Consider 
also the $\sigma$-algebra that informs about the subject's 
complete disease state history until $t$, say $\mathcal{H}_i(t) 
= \{ X_i(u), \E 0 \leq u \leq t \}$. The multistate structure 
can be completely characterized by the transition hazard 
between subsequent states $\{ \ell, m \} \in \mathcal{R}, \ell 
\neq m$, defined as the $i$th subject's instantaneous cause-specific 
hazard per time unit of going from the state $\ell$ to the state 
$m$. Formally, 
\vspace{0.1cm}
\begin{equation}\label{eq:1}
h_i^\lm \{ t \mid \bz_i, \mathcal{H}_i(t) \} = 
  \lim_{\mathrm{d}t \to 0} \frac{\Pr \E \{ X_i(t + \mathrm{d}t) = m \mid 
  X_i(t) = \ell; \E \bz_i, \mathcal{H}_i(t)\} }{\mathrm{d}t},~~t > 0, \\[2.5pt]  
\end{equation}
so that the above hazard-based measure is conditioned on: the 
current time $t$, a $p$-dimensional vector of baseline prognostic 
covariates (possibly different for each transition), $\bz_i = 
(z_{1i}, \dotsc, \E z_{pi})^\red{\top}$, and the historical 
information up to immediately before time $t$, $\mathcal{H}_i(t)$. 
As the other key element within the multistate modeling framework, 
one can define the transition probability between any pair of states, 
even if they are not subsequent. This is postulated as the $i$th 
subject's probability of being in state $m$ at time $t$ conditional 
on occupying the state $\ell$ at time $s$, with $s < t$, along with 
the subject's characteristics $\bz_i$ and $\mathcal{H}_i(s)$:
\begin{equation}\label{eq:2}
P_i^\lm (s, t) = \Pr \{ X_i(t) = m \mid X_i(s) = \ell, \E 
  \bz_i, \mathcal{H}_i(s) \},~~ s < t. \\[1.0pt]  
\end{equation}
For estimation purposes, the underlying multistate process 
is typically assumed to be Markovian conditional on 
transition-specific covariates, so the future and the 
past history are independent given the present. In 
practice, the effect of $\mathcal{H}_i(t)$ disappears, 
which greatly simplifies the parameter estimation procedure. 
Under a nonparametric Markovian framework, the product-limit 
relation given by the Aalen-Johansen (AJ) estimator (Aalen 
\& Johansen, 1978)~provides an easy-to-compute way to 
translate the $\ell \to m$ transition intensities at any $t$ 
into consistent estimates for the transition probabilities. 
If using a Cox semiparametric model (Cox, 1972)~to introduce 
prognostic covariates $\bz_i$ in transition hazards, the 
AJ estimator can also be applied to estimate the corresponding 
transition probabilities (Andersen et al., 1991).~In contrast, 
when departing from the Markov condition, explicit formulas 
for the transition probabilities may be challenging to derive 
(Meira-Machado et al., 2006; Andersen \& Perme, 2008; Titman, 
2015).~The Markov assumption is, nevertheless, a quite 
restrictive hypothesis whose appropriateness needs to be 
properly verified to prevent misleading results. 

The whole multistate process can be modeled by separately 
fitting a semiparametric Cox model for each transition hazard, 
which allows researchers to incorporate the effect of prognostic 
covariates. Under this framework, a subject's past history can 
be easily considered in the $\ell \to m$ transition by including 
some functional form $f(\cdot)$ of the time $t_{\ell i}$ at 
which the current state $\ell$ is reached. For conciseness, 
$f(\cdot)$ is assumed to be the identity, although nothing 
prevents us from employing a more suitable functional form, 
such as high-order polynomials or splines. The subject's 
instantaneous hazard is written as\vspace{0.10cm}
\begin{equation}\label{eq:3}
h_i^\lm ( t \mid \bz_i, t_{\ell i} ) = 
  h_\lar{0}^\lm(t)
  \exp \bigl\{ (\bds{\B}^\lm)^\red{\top} \E \bz_i 
     + \G^\lm \E t_{\ell i} \bigl\}, 
\end{equation}
where $h_\lar{0}^\lm(t)$ is an arbitrary nonnegative 
transition-specific baseline hazard function permitted 
to vary over time, $\bds{\B}^\lm = (\B_{z_1}^\lm, 
\dotsc, \E \B_{z_p}^\lm)^\red{\top}$ is a vector of 
$p$ unknown regression coefficients corresponding to 
$\bz_i$, and $\G^{\lm}$ is an unknown coefficient 
related to $t_{\ell i}$; whenever the Markovian rule 
holds, $\G^\lm = 0$. We therefore have a semi-Markov 
multistate model in which the subject's historical 
effects are accounted for via the covariate describing 
the time of entry into the current state $t_{\ell i}$. 
It is worth mentioning that this latter time does not 
entail a new time scale, but simply incorporates a 
time-of-entry covariate whose value is automatically 
established for those subjects who reach the $\ell$ 
state. As another point, the semiparametric Cox model 
assumes both the proportional hazards hypothesis (i.e., 
constant effect of a given covariate on the hazard over 
the follow-up period) and a log-linear association 
between the transition hazard and covariate. 
Consequently, both hypotheses should be checked.

Estimates for $\bds{\B}^\lm$ and $\G^\lm$ are found by 
maximizing the Cox partial log-likelihood function for 
the transition at hand, whereas the Breslow estimator 
(Breslow, 1972) provides a nonparametric estimate for 
the corresponding cumulative baseline hazard function. 
Here, the \cx{mstate}~package~(Putter et al., 2007)~from 
the \textsf{R} environment is used to perform the 
multistate model fitting, with each transition-specific 
Cox model being fitted independently; this entails a more 
efficient method than maximizing the full joint likelihood 
function over all transitions~(Ieva et al., 2017).~Further 
details regarding parameter estimation can be consulted, 
for example, in de Wreede et al. (2010).
\subsection{Testing the Markov hypothesis}
We describe two existing methods for testing the Markov assumption 
when modeling transition intensities under a semi-Markov Cox-based 
framework. On the one hand, a general procedure is applied for 
revealing discrepancies with respect to the Markov property in any 
given Cox-like transition hazard. This procedure involves a global 
test score, computed in a summarized manner from a family of local 
log-rank tests whose $p$-values are obtained via subsampling 
techniques. On the other hand, as a more particular alternative, one 
may include a covariate which collects certain aspect of the past 
history. This is a useful way to proceed when such an aspect may be 
held liable for departures from the Markovian behavior. 
\subsubsection{Likelihood ratio test to detect any relevant aspect from past history}
Consider an $\ell \to m$ transition hazard that has been estimated 
by means of a Cox regression model. To test the Markov assumption, 
the simplest procedure consists of capturing the subject's past 
history by including the corresponding time of entry $t_{\ell i}$ 
into the current state $\ell$. This provides a rapid and easy manner 
for checking the Markov hypothesis through a likelihood ratio test 
of $\G^\lm = 0$, at a given significance level $\alpha$. Of note, 
an alternative interpretation of such an entry time could be derived 
by assuming the transformation $t_{\ell i} = t - (t - t_{\ell i})$. 
This yields (\ref{eq:3}) to be rewritten as \vspace{0cm}
\begin{equation*}
h_i^\lm ( t \mid \bz_i, t_{\ell i} )  = 
   h_\lar{0}^\lm(t) \exp ( \G^\lm \E t )
  \exp \big\{ (\bds{\B}^\lm)^\red{\top} \E \bz_i 
  - \G_\lar{\mathcal{H}}^\lm \bigl( t - t_{\ell i} \bigr) \big\}, \\
\end{equation*}
providing a direct association between the hazard and both 
the chronological time $t$, collected via the baseline hazard 
function, and the sojourn time in the state $\ell$, i.e., 
$t - t_\ell$. The latter, however, has a regression coefficient 
of opposite sign to $t_\ell$, so the association with the 
transition hazard should now be interpreted in the reverse 
sense. Throughout the article, the results will be presented 
relating to the effect of $t_\ell$, unless otherwise specified.

As in the Markov environment, an advantageous characteristic 
of the Cox semi-Markov multistate model relies on its ease 
of implementation with standard software. Obviously, 
alternative choices for tackling historical follow-up could 
be considered (e.g., the number of states that a subject 
has gone through prior to entering into the state $\ell$). 
The choice of one or the other depends on the degree of 
understanding about the process.
\subsubsection{Global score test for the Markov assumption}
Taking the so-called landmark techniques (de Uña-Álvarez \&  
Meira-Machado, 2015; Putter \& Spitoni, 2018) as a starting 
point, Titman and Putter (2022)~have recently proposed a 
Cox-based global test for assessing the Markov assumption 
conditional on transition-specific covariates. Let $j$ 
be a qualifying state for the $\ell \to m$ transition, 
and $Y_i(t)$ be the subject's at-risk indicator within 
the multistate process. For a given landmark time $s$, 
consider the subsamples $\mathcal{S} = \{ i: X_i(s) = j, 
Y_i(s) = 1\}$ and $\mathcal{S}^c = \{ i: X_i(s) \neq j, 
Y_i(s) = 1\}$. We can test the Markov property by focusing 
on the $i$th subject's membership in $\mathcal{S}$. Under 
Markovian conditions, transition probabilities are 
straightforwardly obtained from transition intensities, 
so differences between $\mathcal{S}$ and $\mathcal{S}^c$ 
can be assessed by merely testing
\begin{equation}\label{eq:4}
\tx{H}_\lar{0} \colon h_i^\lm 
\{ t \mid \bz_i, X_i(s) = j \} = h_i^\lm \{ t \mid \bz_i, X_i(s) 
\neq j \}~\tx{for}~t \geq s,~\tx{against any other alternative}. 
\end{equation}
Testing this hypothesis is carried out via a global score test 
constructed for Cox-based transition intensities. Specifically, 
let $\delta_i^{(j)}(s) = I\{ X_i(s) = j \}$ be the membership 
indicator for the $j$ state. Let also $Y_{i \ell}(t) = I\{ 
X_i(t^{-}) = \ell \} Y_i(t)$ be the at-risk indicator for the 
$\ell \to m$ transition, and $N_i^{\ell m}(t)$ the counting process 
reporting the number of stochastic movements between $\ell$ and $m$ 
up to time $t$. Under a Cox-Markov multistate model, $\tx{H}_\lar{0}$ 
can be tested via the log-rank statistic 
\begin{equation}\label{eq:5}
U^{\ell m (j)}(s, \bds{\B}^\lm )  = 
 \sum_{i=1}^n \int_s^\tau \bigg\{ \delta_i^{(j)}(s) - 
 \frac{\sum_{r=1}^n \delta_r^{(j)}(s) Y_{r \ell}(t) \exp \{ (\bds{\B}^\lm)^\red{\top} \E \bz_r \} }
  {\sum_{r=1}^n Y_{r \ell}(t) \exp \{ (\bds{\B}^\lm)^\red{\top} \E \bz_r \}} \bigg\} \E 
  \mathrm{d}N_i^{\ell m}(t),
  ~~t>s,
\end{equation}
which acts as a score statistic for $\delta_i^{(j)}(s)$ when 
replacing $\bds{\B}^\lm$ with its maximum partial likelihood 
estimate, $\bds{\hat{\B}}{}^\lm$. The first term from the 
difference in (\ref{eq:5}) checks whether a subject is actually 
in the $j$ state at time $s$ or not, whereas the second term is 
the probability that a subject transitioning at time $t$ was 
previously in $j$ at time $s$, conditional on the set of subjects 
in $\ell$ at time $t$. It is essentially a testing procedure 
involving the comparison of the future rate for the possible 
$\ell \to m$ transitions across the different groups defined 
by the $j$ state occupied at time $s$. Thus, $U^{\ell m (j)}(s, 
\bds{\hat{\B}}{}^\lm )$ will differ from zero when the probability 
to reach the $\ell$ state differs provided that $X_i(s)~\neq~j$. 

Because time $s$ is allowed to vary over a period $[t_0, t_\tx{max}] 
\subset [0, \tau]$, a sequence of tests like (\ref{eq:4}) can be 
longitudinally conducted over a grid $\{s_1, \ldots, s_\tx{L} \} 
\in [t_0, t_\tx{max}]$, adequately chosen to avoid intervals with 
a small number of subjects. A log-rank test statistic vector 
$\{ U^{\ell m (j)}(s_1, \bds{\hat{\B}}{}^\lm ), \ldots, 
U^{\ell m (j)}(s_\tx{L}, \bds{\hat{\B}}{}^\lm ) \}$ 
is then obtained, which is proven to be asymptotically independent 
and zero-mean normal under $\tx{H}_\lar{0}$. Thus, the standardized 
process
\begin{equation*}
\bar{U}^{\ell m (j)}(s, \bds{\hat{\B}}{}^\lm ) = 
U^{\ell m(j)}(s, \bds{\hat{\B}}{}^\lm ) / 
[ \tx{Var}^{\ell m(j)}(s, \bds{\hat{\B}}{}^\lm ) ]^{1/2}, 
~ s \in [t_0, t_\tx{max}], 
\end{equation*}
converges to a standard normal distribution, approximated by 
repeatedly generating a large number of resamples (a minimum 
of 1000 is recommended to perform the test) via wild 
bootstrapping methods~(Mammen, 1993).~Three summary test 
statistics are compared according to different assumptions 
regarding how the past-history effect is considered when 
computing (\ref{eq:5}). The absolute mean, $\int_{t_0}^{t_\tx{max}} 
| \bar{U}^{\ell m (j)}(s, \bds{\hat{\B}}{}^\lm ) | 
\mathrm{d}s$, assumes identical importance for all landmark 
times. The weighted absolute mean, $\int_{t_0}^{t_\tx{max}} | 
w(s) \bar{U}^{\ell m (j)}(s, \bds{\hat{\B}}{}^\lm )| 
\mathrm{d}s$ involves weights $w(s)$ chosen to overweight the 
influence of times in which it is more likely to detect Markovian 
departures. The absolute maximum $\tx{sup}_{s \in [t_0, 
t_\tx{max} ]}| \bar{U}^{\ell m (j)}(s, \bds{\hat{\B}}{}^\lm)|$ 
only accounts for the time with the most important deviation 
from the Markov property. By accurately combining the $p$-values 
derived from the sequence of standardized statistics, a global 
$p$-value for testing the Markov property in the selected 
transition can then be returned; the lower the $p$-value, the 
stronger the suggestion that the Markovian assumption is not 
tenable. Further, when handling multiple interconnected states, 
a given landmark $s$ can be associated with two or more qualifying 
states $\{j_1, j_2, \ldots \}$ from which the $\ell \to m$ transition 
is reachable, yielding $\{ \bar{U}^{\ell m (j_1)}(s, 
\bds{\hat{\B}}{}^\lm ), \bar{U}^{\ell m (j_2)}(s, 
\bds{\hat{\B}}{}^\lm ), \ldots \}$. Here, the Markov 
assumption can be globally evaluated by means of an overall 
chi-squared test statistic $T^\lar{\ell m (j)}(s, 
\bds{\hat{\B}}^\lm)$ obtained from $\sum \{ 
\bar{U}^{\ell m(j)}(s, \bds{\hat{\B}}{}^\lm )\}^2$, 
so that an overall $p$-value can be given. 

Complete details on the above testing procedure, as well as 
on how to choose summaries and weights, are given in the 
aforementioned Titman and Putter's article. The test is 
implemented in the \textsf{R} package \cx{mstate} 
through the built-in function termed \cx{MarkovTest}.
\subsection{Formulation of the two Cox-based multistate modeling approaches}
From the Cox-based transition hazard postulated in (\ref{eq:3}), 
the multistate modeling approaches M1 and M2 are considered for 
analyzing event history data from disjoint cohorts. The motivation 
behind both approaches is to account for possible inter-cohort 
heterogeneity in the target population, which ultimately may allow 
us to infer cohort-specific medical conclusions. Let us consider a 
heterogeneous target population that can be split into $G > 1$ 
cohorts of subjects. The $i$th subject's cohort membership, say 
$g = 1, \ldots, G$, is accounted for by $G - 1$ binary indicators, 
namely $\{ c_{gi}, ~ g \geq 2 \}$, where $c_{gi} = 1$ if the subject 
is actually in cohort $g \in \{2, \ldots, G\}$, or $c_{gi} = 0$ 
otherwise. For the particular case $c_{2i} = \ldots = c_{\red{G}i} = 
0$, the subject is part of the first cohort, $g = 1$. The way 
in which the information about cohorts is accounted for yields 
the approaches that are detailed next.
\subsubsection{M1 Approach: Inclusion of a cohort effect and its interaction with a covariate}
The M1~approach accounts for the multicohort nature of the data 
by simultaneously adding two terms in a Cox transition hazard. To 
begin with, we incorporate a collection of $G - 1$ cohort binary 
regressors as fixed effects, along with their corresponding 
regression parameters. However, rather than considering the cohort 
effect in isolation, one may be more interested in assessing how 
the strength of association between a generic transition hazard 
and a given prognostic covariate at baseline is modulated by the 
effect of a specific cohort. Hence, attention can be restricted to 
exploring whether the relationship between the $\ell \to m$ 
transition-specific hazard and a particular time-fixed predictor 
$z_q$ from vector $\bz$, with $q \in \{1, \ldots, p\}$, may 
vary across the distinct cohorts. When assessing this potential 
dependency, the fixed-effects M1~approach advocates for also 
including an interaction term between the cohort $g$ and the 
baseline prognostic covariate $z_q$. This yields the following 
Cox proportional hazards regression model whenever a given subject 
$i$ is in the cohort $g$:
\begin{equation}\label{eq:6}
h_i^\lm \E ( t \mid \bz_i, t_{\ell i}, c_{gi} = 1 ) =
h_\lar{0}^\lm(t) 
 \exp \big\{ (\bds{\B}^\lm)^\red{\top} \E \bz_i
   + \G^\lm \E t_{\ell i}
   + \eta_{g}^\lm 
   + \eta_{g \pd z_q}^\lm \E z_{q i} \big\}. 
\end{equation}
To quantify the effect of parameters from the above Cox-based 
regression, the transition-specific hazard ratio is typically 
used. This quantifies the change in $h_i^\lm(\cdot)$ 
for an increase of $\Delta$ units in the $q$th covariate, 
while conditioning on the cohort. Assuming $z_q$ as continuous 
in the remainder of the article, the corresponding conditional 
hazard ratio is defined at any time $t$ by \\[-2.5pt]
\begin{equation}\label{eq:7}
\tx{HR}_{z_q \mid g}^\lm = 
 \exp \big\{ \big( \B_{z_q}^\lm
    + \eta_{g \pd z_q}^\lm \big) \Delta \E \big\}. \\[5pt]
\end{equation}
A question of interest lies then in testing whether, conditional 
on the cohort $g$, the hazard ratio associated to a $\Delta$-$\E$unit 
increase in $z_q$ changes significantly. We then perform the test
\begin{equation}\label{eq:8}
\tx{H}_\lar{0} \colon 
\eta_{\red{2} \pd z_q}^\lm = \ldots = 
  \eta_{\red{G} \pd z_q}^\lm = 0 ~~\tx{against}~~ 
\tx{H}_\tx{A} \colon \exists ~g \in \{2, \ldots, G \} ~\tx{such that}~ 
  \eta_{g \pd z_q}^\lm \neq 0, \\[5pt]
\end{equation}
which can be conducted using a likelihood ratio test.

On the other hand, we can compare the transition hazard of 
a subject from the cohort $g \in \{2, \ldots, \E G\}$ with 
respect to the transition hazard attributable to the 
first-cohort membership, while keeping the $q$th covariate 
fixed at some value $z_q$ across cohorts. The corresponding 
hazard ratio takes the form 
\begin{equation}\label{eq:9}
\tx{HR}_{g \mid z_q}^\lm = \exp \big( \eta_{g}^\lm + 
   \eta_{g \pd z_q}^\lm z_q \big),\\[2.5pt]
\end{equation}
through which the test (\ref{eq:8}) can be once again performed.
\subsubsection{M2 Approach: Cohort stratification, allowing the cohort to interact with a covariate}
For our M2~approach, we assume a stratum-cohort Cox model, with the
cohort serving as the stratification variable. Each cohort's underlying 
risk is then characterized by its own baseline hazard function, while 
still sharing the estimated values for the regression parameters. 
Consequently, the proportional hazards hypothesis may be said to hold 
within each cohort, but not across the cohorts. Under the M2 approach, 
the $\ell \to m$ transition hazard for a subject $i$ belonging to the 
cohort $g = 1, \ldots, G$ is given by 
\begin{equation}\label{eq:10}
h_{i}^\lm \E ( t \mid \bz_i, t_{\ell i}, c_{gi} = 1 ) =
h_{0\lar{g}}^\lm \big(  t \big)
 \exp \big\{ (\bds{\B}^\lm)^\red{\top} \E \bz_i
   + \G^\lm \E t_{\ell i}
   + \eta_{g \pd z_q}^\lm z_{q i}  \big\},\\[2.5pt]
\end{equation}
where $h_\lar{0g}^\lm(t)$ is the baseline hazard function specific 
to the $g$th cohort. Note that the isolated cohort effect has now 
vanished, even though the model allows for estimating the way in 
which some covariate of interest $z_q$ is modulated within a given 
cohort. Moreover, the hazard ratio associated with a 
$\Delta$-$\E$unit increase in $z_q$ is again conditioned on 
the cohort:
\begin{equation}\label{eq:11}
\tx{HR}_{z_q \cx{|} g}^\lm = 
 \exp \big( \eta_{g \pd z_q}^\lm \Delta \big). \\[2.5pt]
\end{equation}
Nonetheless, under M2~approach, comparison of transition hazards 
from subjects belonging to different cohorts with identical $z_q$ 
cannot be done via a simple parameter. Rather, the expression for 
the hazard ratio becomes time-dependent in that case:
\begin{equation*}
\tx{HR}_{z_q \cx{|} g}^\lm (t) = 
    \frac{h_{0g}^\lm (t)}{h_{01}^\lm (t)}
 \exp \big\{ \big( \eta_{g \pd z_q}^{\ell m} - \eta_{1 \pd z_q}^{\ell m} \big) \E z_q \big\}.
\end{equation*}
\section{Analysis of the empirical data from three \mbox{COVID-19} waves}
Based on Figure~1, the analysis of the \textsf{div3W}~data 
is concerned with attaining a better characterization of the 
distinct \mbox{COVID-19} waves. For covariate selection at 
a given transition, the main prognostic covariates to be 
possibly included are those of Table~1, along with the 
time of entry into the current state for non-Markovian 
transitions and the wave covariate within the M1 approach. 
Moreover, since interaction terms between wave and any 
covariate are allowed for M1 and M2 approaches, the 
quantitative prognostic covariates are centered around 
their means to preserve meaningful interpretations of 
their main effects. Thus, leaving aside the inclusion 
potential interactions, the main covariate effects to be 
considered are listed in Table~2.
\vspace{-0.2cm}
\begin{table}[ht]
\centering\setlength{\aboverulesep}{0pt}\setlength{\belowrulesep}{0pt}
\setlength{\extrarowheight}{1pt}
\caption{Potential baseline covariates to be included for modeling 
the $\ell \to m$ transition hazard from the \textsf{div3W}~data, under 
M1 and M2 approaches.}
\begin{tabular}{lc}
\rowcolor{gray!40}
\tb{Covariate notation} & ~~\tb{Meaning}                     \\
\midrule[\heavyrulewidth]
sex (female)            & ~~Indicator for female                  \\
age (years)             & ~~Centered age                          \\
safi (units)            & ~~Centered safi value                   \\
$t_\ell$ (days)         & ~~Time of entry into the $\ell$ state    \\
$g$ (wave)              & ~~Wave factor, $g = 1,2,3$ (only for M1) \\[1pt]
\bottomrule
\end{tabular}\\[-8pt]
\end{table}    
\par In a second phase, a minimum number of individuals per 
transition is checked in order to achieve reliable parameter 
estimates. In practice, this compels us to ensure a sufficiently 
large sample size for the number of parameters to be estimated. 
We have followed Vittinghoff \& McCulloch (2007),~who recommend 
that medical researchers consider at least five subjects per 
predictor within the Cox modeling setting. Using this general 
rule of thumb in our data, a given transition hazard will only 
consider the baseline hazard of occurrence whenever $n \leq 5$ 
in any of the three waves. If this threshold is surpassed in all 
waves, the transition hazard will accommodate as many prognostic 
covariates as possible from Table~2, starting with the inclusion 
of sex and age covariates as potential cofounders; they are often 
easy to measure, generally having few or no missing values.

The distinct transition hazards will be estimated under the M1 
and M2 approaches, while examining the Markov property and the 
wave effect accommodation. Throughout, a significance level of 
0.05 is adopted to examine the relevance of our results.
\subsection{Application of the global test for the key transitions}
The Markov property, conditional on transition-specific covariates, 
is only assessed in certain transitions described by our empirical 
data. In that respect, note that transitions starting from the NSP 
state are inherently Markovian, and therefore excluded from any 
assessment. Moreover, in order to ensure worthwhile estimates, 
transitions involving $n \leq 5$ subjects in any wave are not 
considered. Finally, the sojourn time in the recovery state has 
been previously set at 2 days, so the Markovian assessment is not 
carried out for trajectories starting or ending at this state. For 
the remaining transitions, a sequence of Cox-based log-rank tests 
is applied over a suitable grid of landmark times for both the M1 
and M2 approaches. Adequately selecting the test grid, however, 
is often not a trivial task: highly unbalanced subsamples may 
arise at early times, while too-small subsample sizes can be 
obtained at the later ones. For each transition, $p$-values from 
different grids are compared to determine appropriate choices 
for the landmarks. This leads us to compute the test statistics 
over the follow-up interval $[t_0 = 1, t_\tx{max} = 10]$ for the 
SP~$\to$~NIMV and SP~$\to$~IMV transitions, while considering 
$[t_0 = 2, t_\tx{max} = 10]$ for the NIMV~$\to$~IMV and 
IMV~$\to$~death transitions. As information provided by 
covariates, the M1 and M2 approaches account for the same 
prognostic information within a given transition, while only 
differing in the way they deal with the wave effect. Further, 
three alternatives for summarizing the resulting vector of test 
statistics $\{ \bar{U}^{\ell m (j)}(s, \bds{\B}^\lm), ~ s \in 
[t_0, t_\tx{max}] \}$ are considered: the mean, the weighted 
mean, and the supremum. In each case, the results are obtained 
from 2000 wild bootstrap resamples.

The $p$-value estimates of the global score test for all the 
transitions analyzed can be found in Sections~A.1$\E$--$\E$A.2 
of Appendix~A (Tables S1$\E$--$\E$S2). For conciseness, Table~3 
of the manuscript shows only the results derived from those 
transitions where the Markov property, conditional on observed 
covariates, is ultimately rejected in any of the \mbox{COVID-19} 
waves: SP~$\to$~NIMV and SP~$\to$~IMV. Particularly, the overall 
$p$-values provide evidence to reject the Markov hypothesis for 
these transitions under the M1 and M2 approaches. Hence, the 
results for both approaches suggest that subject's past history 
has a determining role when moving from SP to any one of the 
two states related to the need for mechanical ventilation, 
whereas these historical effects are not found to have an 
influence in the in the NIMV~$\to$~IMV and IMV~$\to$~death 
transitions for any of the waves. An intuitive explanation 
of this pattern is as follows. Considering that all the 
hospitalized subjects under study are, by definition, critically 
ill individuals with \mbox{COVID-19}, departures from the Markov 
assumption must be found in differences between baseline prognostic 
characteristics from subjects that were hospitalized without SP 
versus those with SP. That is, baseline covariates would become 
especially relevant in the first days following hospital admission. 
Nevertheless, as the multistate process moves forward and subjects 
reach more critical states, the possible past-history prognostic 
role vanishes progressively. In some sense, the health risk would 
tend to be similar for subjects occupying states far from hospital 
admission, so the NIMV~$\to$~IMV and IMV~$\to$~death transitions 
would become essentially Markovian conditional on observed baseline 
covariates.
\par
\begin{table}[ht!]
\centering\setlength{\aboverulesep}{0pt}\setlength{\belowrulesep}{0pt}
\setlength{\extrarowheight}{0pt}
\caption{Results of the global score test for the assessment of the Markov 
property conditional on covariates, using both the M1 and M2 approaches 
to analyze the \mbox{\textsf{div3W}} data. The presentation of results is here 
restricted to the non-Markovian transitions, for which the global test provides 
$p$-value estimates both at each qualifying state and in overall terms, while 
considering three distinct summaries from families of log-rank statistics: 
unweighted mean (UM), weighted mean (WM), and the supremum (S). 
\label{T2}} 
\begin{tabular}{clcrrccr}
\rowcolor{gray!40}
\rule{0pt}{12pt}  
     &                          &         &      \multicolumn{4}{c}{\tb{Qualifying states}}  &  \\[2.5pt]
\rulefiller\cmidrule[0.275ex](lr){4-7} 
\rowcolor{gray!40}
\rule{0pt}{12pt}   
\tb{Approach}  & \tb{~~~~Transition} & \tb{Rule} & \multicolumn{1}{c}{~~\tb{NSP}} & \tb{SP}~~ & \tb{NIMV} & \tb{IMV} & \tb{Overall}  \\
\midrule[0.275ex]
\rule{0pt}{12pt}  
 M1 &    SP $\to$ NIMV    &  UM    &  $<\E$0.001  &   $<\E$0.001 &  \CT    &  \CT   & $<\E$0.001  \\
    &                     &  WM    &  $<\E$0.001  &   $<\E$0.001 &  \CT    &  \CT   & $<\E$0.001  \\
    &                     &  S     &    0.008     &   $<\E$0.001 &  \CT    &  \CT   & $<\E$0.001  \\  
    &    SP $\to$ IMV     &  UM    &    0.007     &        0.002 &  \CT    &  \CT   &      0.003  \\
    &                     &  WM    &  $<\E$0.001  &   $<\E$0.001 &  \CT    &  \CT   & $<\E$0.001  \\
    &                     &  S     &    0.056     &        0.008 &  \CT    &  \CT   &      0.008  \\
\rowcolor{gray!20}
\rule{0pt}{12pt}  
 M2 &   SP $\to$ NIMV    &   UM   &  $<\E$0.001  &   $<\E$0.001  &  \CT    &  \CT   &  $<\E$0.001 \\
\rowcolor{gray!20}
    &                    &   WM   &  $<\E$0.001  &   $<\E$0.001  &  \CT    &  \CT   &  $<\E$0.001 \\
\rowcolor{gray!20}     
    &                    &   S    &    0.011     &   $<\E$0.001  &  \CT    &  \CT   &  $<\E$0.001 \\
\rowcolor{gray!20}
    &   SP $\to$ IMV     &   UM   &    0.002     &   $<\E$0.001  &  \CT    &  \CT   &  $<\E$0.001 \\
\rowcolor{gray!20}
    &                    &   WM   &  $<\E$0.001  &   $<\E$0.001  &  \CT    &  \CT   &  $<\E$0.001 \\
\rowcolor{gray!20}
    &                    &   S    &    0.003     &   $<\E$0.001  &  \CT    &  \CT   &  $<\E$0.001  
\end{tabular}
\end{table}
\subsection{Parameter estimates within the three-wave semi-Markov multistate model}
Each of the transition hazards from the \textsf{div3W} data is 
modeled by separately fitting a Cox semiparametric model, under 
both the M1 and the M2 approaches. Whenever a given transition 
hazard involves a sufficiently large number of individuals in 
each of the three waves, all the mentioned covariates in 
Table~2 are included in the modeling procedure. Moreover, the 
significance of the different interactions between any of 
these covariates and the wave effect are sequentially checked 
via the corresponding bivariate test. For the sake of parsimony, 
in each transition hazard we only keep the interaction term 
with the smallest $p$-value, whenever $p$-value$\E < 0.05$ is 
obtained. Table~4 provides the distinct covariates considered 
when modeling each transition hazard, while indicating the 
$p$-value associated to the significant interaction term, if 
any, that is included. Furthermore, for those transitions in 
which the Markov assumption does not hold, i.e. SP~$\to$~NIMV 
and SP~$\to$~IMV, the time to entry into SP, say \tsp, is 
included as an additional covariate. To assess the proper 
functional form $f(\cdot)$ trough which~\tsp~should be 
incorporated, we can compare the performance of this covariate 
under two distinct shapes. Initially, one can assume the 
traditional linear assumption for this covariate on the 
log-hazard scale, i.e. $f(\cdot) = \tx{Id}(\cdot)$. As an 
alternative, this linear assumption can be relaxed by embedding 
\tsp~into a natural cubic spline basis. The two resulting fits 
can be then compared using a likelihood ratio test that informs 
us about possible nonlinear effects. This comparison is provided 
for both approaches in Section~B.1 of Appendix~B (Figures 
S1$\E$--$\E$S2), allowing us to conclude that no particular 
transformation is needed for the inclusion of~\tsp. The 
transition-specific point and 95\% confidence interval (CI) 
estimates are provided in Section~B.2 of Appendix~B (Tables 
S3$\E$--$\E$S4), and the corresponding hazard ratios (where 
each covariate is related to a prespecified $\Delta$-unit 
increment) are found in Section~B.3 (Tables S5$\E$--$\E$S6). 
Furthermore, for a given transition-specific covariate, the 
proportional hazards assumption is graphically checked using 
a class of tests statistics based on cumulative sums of 
martingale-based residuals (Lin et al., 1993). Briefly, this 
conforms a multiparameter stochastic process that can be 
approximated by a zero-mean Gaussian process under the 
assumed model, so the observed pattern is compared with a 
number of simulated ones from the null distribution. Further, 
we can compute the Kolmogorov-type supremum statistic test 
based on a large number of realizations, reporting an 
estimate for the $p$-value of the test. The main advantage 
of this omnibus test, implemented in the \textsf{R} software 
by Martinussen \& Scheike (2006), relies on the fact that 
no particular shape needs to be assumed when testing for 
lack of fit regarding a given covariate. For each transition 
and covariate, graphical plots of the score process over 
follow-up time are given in in Section~B.4 (Figures 
S3$\E$--$\E$S22). In most transitions, the PH assumption 
is fulfilled, even though departures from proportionality 
are obtained for the safi and the cohort covariate under 
the M1 approach. Generally, these deviations are corrected 
for the most flexible approach M2.
\begin{table}[ht!]
\centering\setlength{\aboverulesep}{0pt}\setlength{\belowrulesep}{0pt}
\setlength{\extrarowheight}{0pt}
\caption{
  Baseline and historical characteristics to be accounted for
  when estimating each of the transition hazards from the 
  \textsf{div3W}~data under the M1 and M2 approaches.
\label{T3} \\[-15pt]}
  \setlength{\tabcolsep}{5pt}
\begin{tabular}{r@{\hs{5pt}}llccc}
\rowcolor{gray!40}
\rule{0pt}{12pt}   
& \multicolumn{1}{l}{\tb{Transition}}  &  \tb{Prognostic covariates}   & $\bds{t_\ell}$ & \tb{Wave (M1)} & \tb{Interaction ($\bds{p}$-values)} \\
\midrule[0.275ex]
\rule{0pt}{12pt}  
(1)  & NSP~$\to$~SP             &  ~~~~sex, age, safi & \CT  & $g$  &  $g \tm \tx{safi}$~\E $(p_\tx{M1}<0.001, p_\tx{M2}=0.002)$ \\
(2)  & NSP~$\to$~Discharge      &  ~~~~sex, age, safi & \CT  & $g$  &                    \CT                                    \\
(3)  & NSP~$\to$~Death          &      \CT            & \CT  & \CT  &                    \CT                                     \\
(4)  & SP~$\to$~Recovery        &  ~~~~sex, age, safi & \CT  & $g$  &                    \CT                                      \\
(5)  & SP~$\to$~NIMV            &  ~~~~sex, age, safi & \tsp & $g$  &  $g \tm \tx{safi}$~\E $(p_\tx{M1}=0.002, p_\tx{M2}=0.001)$  \\
(6)  & SP~$\to$~IMV             &  ~~~~sex, age, safi & \tsp & $g$  &  $g \tm \tx{safi}$~\E $(p_\tx{M1}=0.041, p_\tx{M2}=0.088)$  \\
(7)  & SP~$\to$~Death           &      \CT            & \CT  & \CT  &                   \CT                                      \\
(8)  & Recovery~$\to$~Discharge &  ~~~~sex, age, safi & \CT  & $g$  &                   \CT                                     \\
(9)  & Recovery~$\to$~Death     &      \CT            & \CT  & \CT  &                    \CT                                      \\
(10) & NIMV~$\to$~Recovery      &  ~~~~sex, age, safi & \CT  & $g$  &                    \CT                                    \\
(11) & NIMV~$\to$~IMV           &  ~~~~sex, age, safi & \CT  & $g$  &  $g \tm \tx{age}$~\E $(p_\tx{M1}=0.016, p_\tx{M2}=0.009)$  \\
(12) & NIMV~$\to$~Death         &      \CT                 & \CT  & \CT  &                  \CT                                       \\
(13) & IMV~$\to$~Recovery       &  ~~~~sex, age, safi & \CT  & $g$  &                   \CT                                     \\
(14) & IMV~$\to$~Death          &  ~~~~sex, age, safi & \CT  & $g$  &                  \CT                                      \\[1pt]
\bottomrule
\end{tabular}\\[-8pt]
\end{table}    
\par Basically, within a given transition, the estimates derived 
from both approaches are found to be very similar. For the M1 and 
M2 approaches, the wave factor exhibits a significant interaction 
with the safi value within the earlier transitions. By contrast, 
in the later transitions, the presence of comorbidities at hospital 
entry do not report behavioral differences across waves: the wave 
membership is shown to only interact significantly with the age 
covariate in the NIMV~$\to$~IMV transition. An explanation for 
this can be found in the fact that the first \mbox{COVID-19} wave 
exhibits a younger subject profile in comparison with the second 
and third waves, wherein the need for supplemental oxygen is 
probably crucial to understanding a subject's transition from NIMV 
to IMV. Instead, upon entering into the IMV state, the subject's 
health status is severe enough that differences in age and 
comorbidity patterns between the waves do not seem to have a 
pivotal influence on the forward transitions.

For brevity, we next focus exclusively on the non-Markovian 
transition between $\ell = \tx{SP}$ and $m = \tx{NIMV}$ states 
to show how graphical tools can be used to analyze the numerical 
output obtained from a given transition. The time of entry into 
SP, \tsp, is incorporated as a historical covariate. Under the 
M1~approach, the SP~$\to$NIMV transition hazard is modeled by 
specifying (\ref{eq:6}) as 
\begin{align*}
\hs{0.15cm} h_i^\lm \E ( t \mid 
   \tx{sex}_i, \tx{age}_i, \tx{safi}_i, 
   t_{\red{\raisebox{0.4ex}{\tx{SP}}}i}, c_{gi} = 1 ) & \notag\\[3.0pt] 
& \hs{-4.00cm}= h_\lar{0}^\lm \big( t \big) 
   \exp \big\{ 
   \B^\lm_\tx{sex} \E \tx{sex}_i
   + \B^\lm_\tx{age} \E \tx{age}_i
   + \B^\lm_\tx{safi} \E \tx{safi}_i   
   + \G^\lm \E t_{\red{\raisebox{0.4ex}{\tx{SP}}}i}
   + \eta_{g}^\lm 
   + \eta_{g \pd \tx{safi}}^\lm \E \tx{safi}_i  \big\}, 
\end{align*}
\vspace{0.05cm}
where the point estimates and 95\% confidence intervals of 
the isolated wave effect are found to suggest an increasing 
propensity to receiving NIMV over the course of the pandemic: 
$\hat{\eta}_{2}^\lm = 0.250~(-0.027, 0.527)$ and 
$\hat{\eta}_{3}^\lm = 0.311~(0.046, 0.586)$. This apparently 
contradictory statement is actually consistent with the 
progressively less strict medical criteria for receiving 
advanced care: the relaxation of this criteria enabled 
access to NIMV for very ill subject profiles who would not 
have been admitted during the first wave.

By means of (\ref{eq:7}), the M1 approach provides the 
corresponding hazard ratio associated with a $\Delta$-$\E$unit 
increase in the safi prognostic covariate:
\vspace{0.15cm}
\begin{equation*}
\tx{HR}_{ \tx{safi} \cx{|} g}^\lm = \begin{cases} 
   \exp \big( \B_{\tx{safi}}^\lm \E \Delta \big) & \tx{ for $g = 1$},  \\[1.5pt]
   \exp \big\{\big( \B_\tx{safi}^\lm
        + \eta_{\red{2} \pd \tx{safi}}^\lm \big) \E \Delta \big\} 
        & \tx{ for $g = 2$},\\[1.5pt]
   \exp \big\{\big( \B_\tx{safi}^\lm 
        + \eta_{\red{3} \pd \tx{safi}}^\lm \big) \E \Delta \big\} 
        & \tx{ for $g = 3$}.\\[1.5pt]
   \end{cases}
\end{equation*}
\vspace{0.05cm}
\par 
\noindent {For the M2~approach, the three waves form the 
natural groups for implementing our stratification. The 
SP~$\to$NIMV transition hazard comes from particularizing 
(\ref{eq:10}) as}
\begin{align*}
\hs{2cm} h_i^\lm \E ( t \mid \tx{sex}_i, \tx{age}_i,  
  \tx{safi}_i, t_{\red{\raisebox{0.4ex}{\tx{SP}}}i}, c_{gi} = 1 ) & \notag\\[3.5pt] 
& \hs{-3cm} = h_{0\lar{g}}^\lm(t) \exp \big(  
   \B^\lm_\tx{sex} \E \tx{sex}_i
   + \B^\lm_\tx{age} \E \tx{age}_i
   + \G^\lm \E t_{\red{\raisebox{0.4ex}{\tx{SP}}}i}
   + \eta_{g \pd \tx{safi}}^\lm \E \tx{safi}_i  \big), 
\end{align*}
\vspace{0.20cm}
where the wave effect itself vanishes from the analysis since it is 
used for stratification. The hazard ratio associated with an increase 
of $\Delta$~units in safi is concretized from (\ref{eq:11}) through
\par
\begin{equation*}
\tx{HR}_{\tx{safi} \cx{|} g}^\lm = 
 \begin{cases} 
 \exp \big( \eta_{\red{1} \pd \tx{safi}}^\lm \E \Delta  \big)
        & \tx{ for $g = 1$},\\[1.5pt]
 \exp \big( \eta_{\red{2} \pd \tx{safi}}^\lm \E \Delta  \big)
        & \tx{ for $g = 2$},\\[1.5pt]
 \exp \big( \eta_{\red{3} \pd \tx{safi}}^\lm \E \Delta  \big)
       & \tx{ for $g = 3$}.
    \end{cases}      
\end{equation*}
\vspace{0.1cm}
\par
Figure~2 summarizes, as a forest plot, the hazard ratio estimates 
and 95\% CI derived for the SP~$\to$~NIMV transition when fitting 
our data under M1 and M2. Interestingly, there is hardly any 
difference between the estimates obtained from both approaches. 
For the baseline prognostic covariates sex (female indicator) 
and age (taking increments of 5 years), the corresponding 95\% CI 
contain the unit value, so we cannot conclude an effect in any 
case. On the other hand, the results from both approaches provide 
above-one hazard ratios associated to $t_\tx{SP}$, with 
$\widehat{\tx{HR}}_\tx{M1} = 1.790$ and $\widehat{\tx{HR}}_\tx{M2} 
= 1.800$, respectively. However, for the sake of interpretation, 
we instead opt for displaying the results in terms of the sojourn 
time in SP, $t - t_\tx{SP}$, yielding $\widehat{\tx{HR}}_\tx{M1} 
= 1/1.790 = 0.559$ and $\widehat{\tx{HR}}_\tx{M2} = 1/1.800 = 
0.556$, with their corresponding 95\% CI clearly below the hazard 
ratio threshold of one unit. There is consequently strong evidence 
for suggesting that the more time a subject passes in SP, the 
lower their risk for moving to NIMV. This makes sense, since very 
ill subjects tend to quickly require ventilatory support and so 
are more correlated with the risk of starting off with NIMV. It 
is worth noting that our approaches account for the interaction 
between the wave factor and the safi value (considering increments 
of $\Delta = 25$ units), thereby allowing us to interpret the 
association between the transition hazard and this covariate 
value, conditional on wave membership. Interestingly, the forest 
plot indicates not only that higher safi values are protective 
with respect transitioning to NIMV, but also shows how the safi 
effect is modulated across the three \mbox{COVID-19} waves. In 
particular, the expected protective impact of higher safi values 
increases progressively across the three waves: $1 > 
\tx{HR}_{\tx{safi} \cx{|} 1}^{\lm} > \tx{HR}_{\tx{safi} 
\cx{|} 2}^{\lm} > \tx{HR}_{\tx{safi} \cx{|} 3}^{\lm}$. That is, 
although a certain increase in safi reduces the hazard for moving 
from SP to NIMV in all waves, a higher protective effect is 
observed among subjects in the second and third waves, which 
paradoxically comprise an increase in subject-specific 
comorbidities in comparison with the first wave. Rather, the 
reason for this difference in behavior between the waves resides 
in a better knowledge of the disease over time, yielding more 
reliable treatment results even when a subject enter the hospital 
with more deteriorated health conditions.
\begin{figure}[!ht]
\begin{center}
\includegraphics[scale = 0.135]{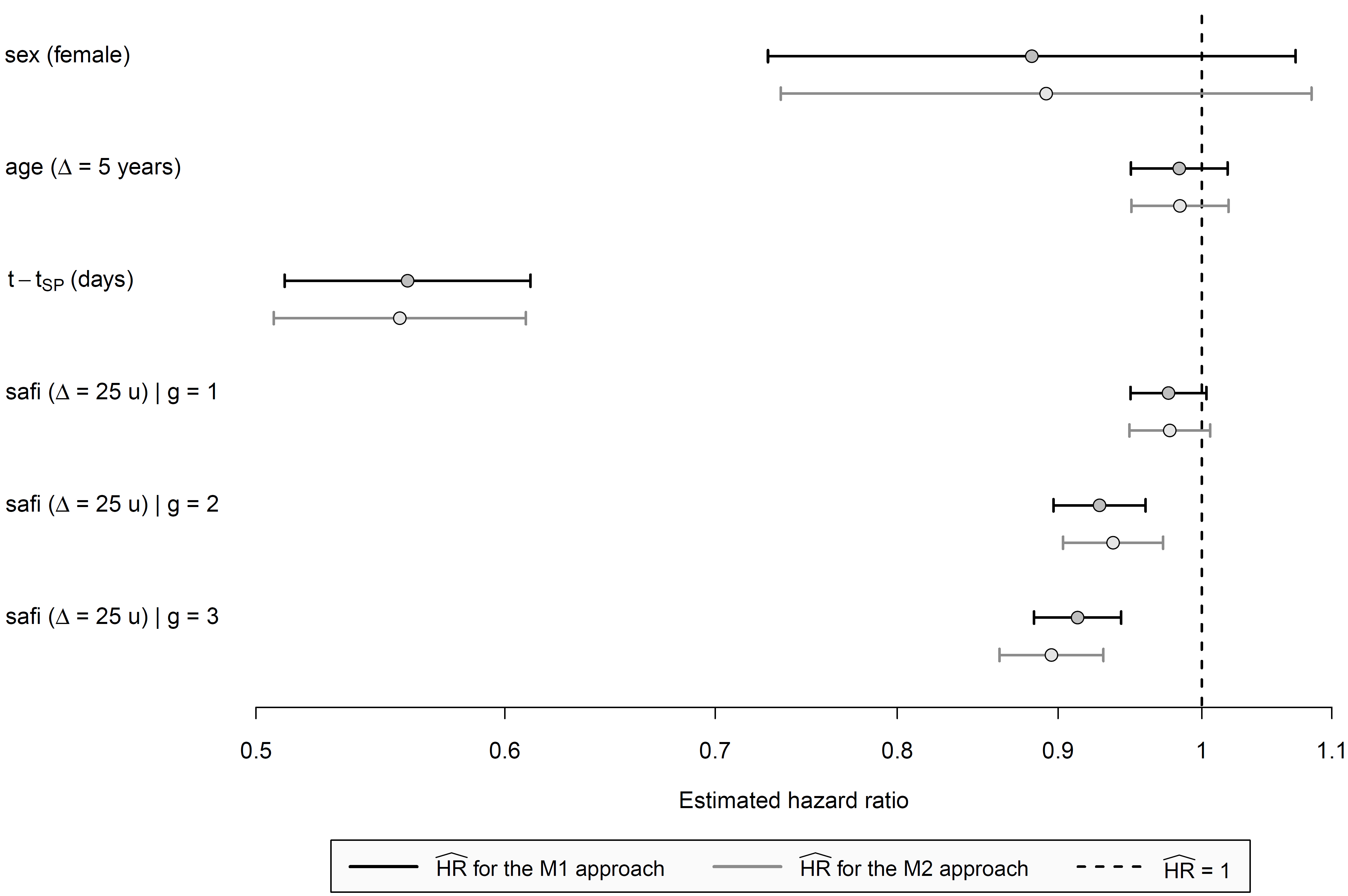}
\end{center}
\caption{Hazard ratio estimates when the M1 and M2 approaches 
are employed to fit the hazard transition 5 (SP $\to$ NIMV) 
from the \textsf{div3W}~data. Standard errors are obtained 
and used to construct the 95\% confidence intervals. \label{F2}}
\end{figure}
\par
As an interesting feature of the M1~approach, it allows for 
comparing the transition-specific hazard associated with belonging 
to either to the second or third wave versus being located in the 
first wave, conditional on a given covariate value. Coming back to 
the SP~$\to$~NIMV transition, we can analyze how the transition hazard 
from SP to NIMV varies across waves according to the specific value 
attained by the safi baseline covariate. This leads (\ref{eq:9}) 
to being concretized as follows:\\
\begin{equation*}
\tx{HR}_{g \cx{|} \tx{safi}}^\lm =
\begin{cases} 
  \exp ( \eta_{\red{2}}^\lm + 
         \eta_{\red{2} \pd \tx{safi}}^\lm \tx{safi}) & 
             \tx{for $g = 2$},\\[1.5pt]
  \exp ( \eta_{\red{3 \E}}^\lm + 
         \eta_{\red{3} \pd \tx{safi}}^\lm \tx{safi}) & 
             \tx{for $g = 3$}.
  \end{cases}
\vspace{0.30cm}  
\end{equation*}
By using the above expression, Figure~3 illustrates how the 
estimated hazard ratios related to the second and third waves 
vary with respect to the first wave, conditioned on attaining 
a given safi value within the range [-100, 81.5]; this 
interval has been obtained from centering (around their mean) 
the original values observed for the safi prognostic 
covariate, the vast majority of which were placed within the 
range [300, 476.2]. Although both $\tx{HR}_{\E 2 \cx{|} 
\tx{safi}}^\lm$ and $\tx{HR}_{\E 3 \cx{|} \tx{safi}}^\lm$ lead 
to above-one quantities when a subject moves between $\ell = 
\tx{SP}$ and $m = \tx{NIMV}$ states, these two quantities 
depict similar decreasing patterns across the considered 
interval for safi. That is, their corresponding protective 
effect as regards going to NIMV becomes higher with the 
safi value. Further, conditional on a given safi, it should 
be noted that subjects who belong to the third wave have a 
higher hazard ratio in comparison to subjects who belong 
to the second wave. That is, the protective effect of 
moving to NIMV is smaller for a third-wave subject. A 
possible explanation for this resides in the capability 
of treating subjects with more severe comorbidities during 
the third wave, who in counterpart would achieve a smaller 
protective effect at a given safi value due to having a more 
deteriorated health status at hospital entry. Hence, though 
the evolution of the disease knowledge cannot be directly 
observed, it does continuously play an underlying role when 
comparing results from different historical periods.
\begin{figure}[ht!]
\begin{center}
\includegraphics[scale = 0.160]{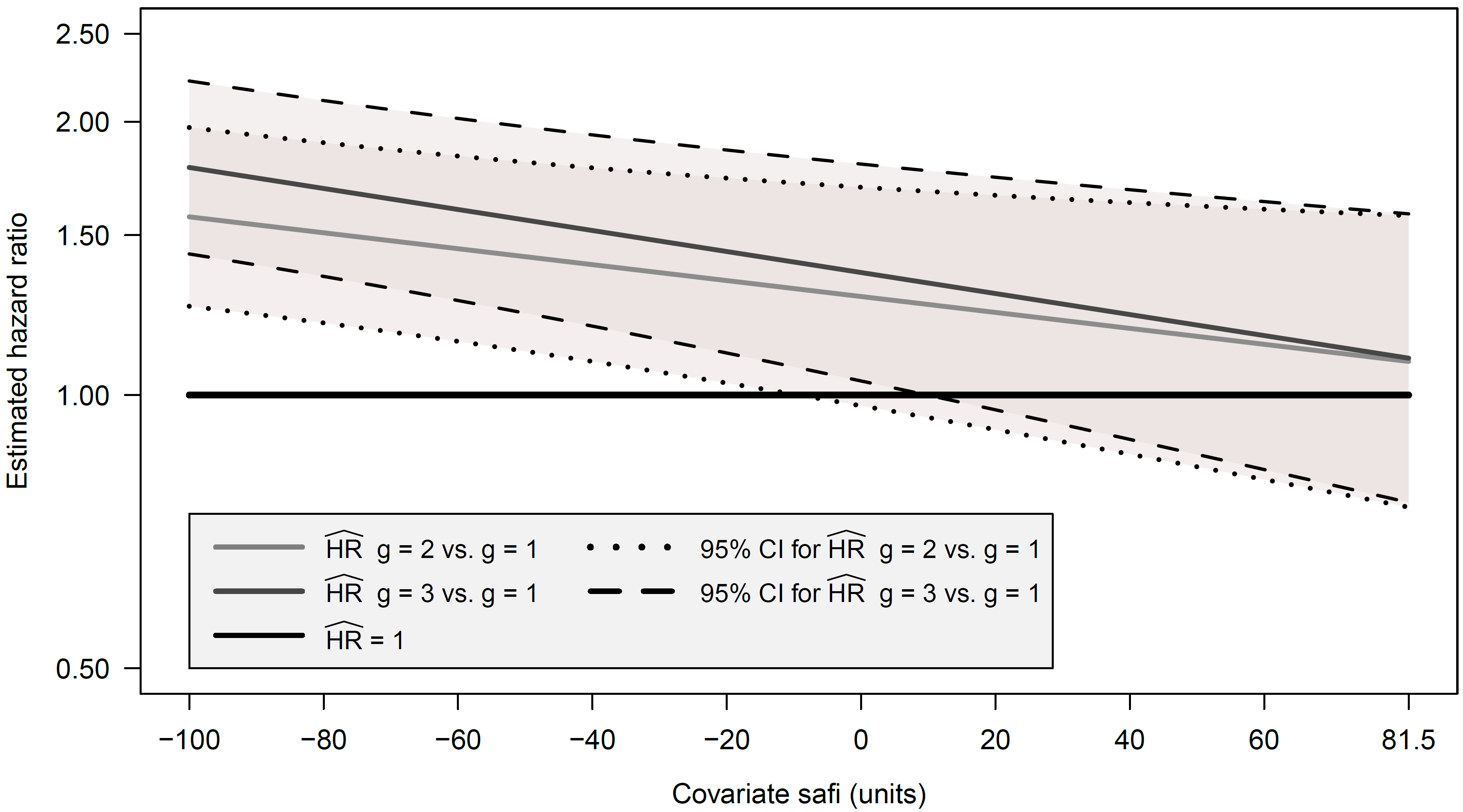}
\end{center}
\caption{Within-wave evolution of the hazard ratio estimates 
when the M1~approach is employed to fit the hazard transition 
5 (SP $\to$ NIMV) from \textsf{div3W}~data, conditioned on 
the safi value. Standard errors are obtained and used to construct 
the 95\% confidence intervals. \label{F3}}
\end{figure}
\vspace{-0.15cm}
\par 
\noindent{Likewise, Figure 4 illustrates the main strength of the 
M2~approach, allowing us to assess how each wave-specific baseline 
hazard evolves as a nuisance function upon entry into the SP state, 
from there moving into one of four potential states: discharge, NIMV, 
IMV, and death. This four-panel figure has a twofold illustrative 
purpose. Firstly, each of the four direct transitions under 
consideration conveys substantial information about the wave-related 
baseline hazard functions themselves. For instance, the panel devoted 
to the SP~$\to$~discharge transition indicates that subjects from the 
first \mbox{COVID-19} wave maintain a relatively constant underlying 
risk over the time interval [0, 40] days, subsequently experiencing a 
sharp increase in the instantaneous risk of being discharged. This 
behavior differs substantially from the other two \mbox{COVID-19} waves, 
which exhibit a gradual increase over time. A possible explanation for 
this discrepancy, as has been noted above, stems from the limitations 
in both the knowledge of the disease and the medical resources during the 
first wave, resulting in longer stays in the SP state for many of the 
hospitalized subjects. As a result, when fixing a reference point within 
the first 40 days of subjects' in-hospital stay, those belonging to the 
first wave have a lower underlying risk of being discharged than those 
in either the second or the third wave, when better \mbox{COVID-19} 
treatments were readily available. Secondly, these plots as a whole 
provide a comprehensive picture of how the flows of subjects moving 
through different transitions are implicitly interrelated. Particularly, 
it is instructive to realize how the hazard of moving from SP to recovery 
increases within a 10-day period, with hardly any subjects moving to the 
NIMV, IMV, and death states after this period. In other words, we may 
infer that there is barely any risk of moving towards more critical states 
once a subject has remained in the SP state for approximately 10 days.}
\begin{figure}[!ht]
\begin{center}
\includegraphics[scale = 0.205]{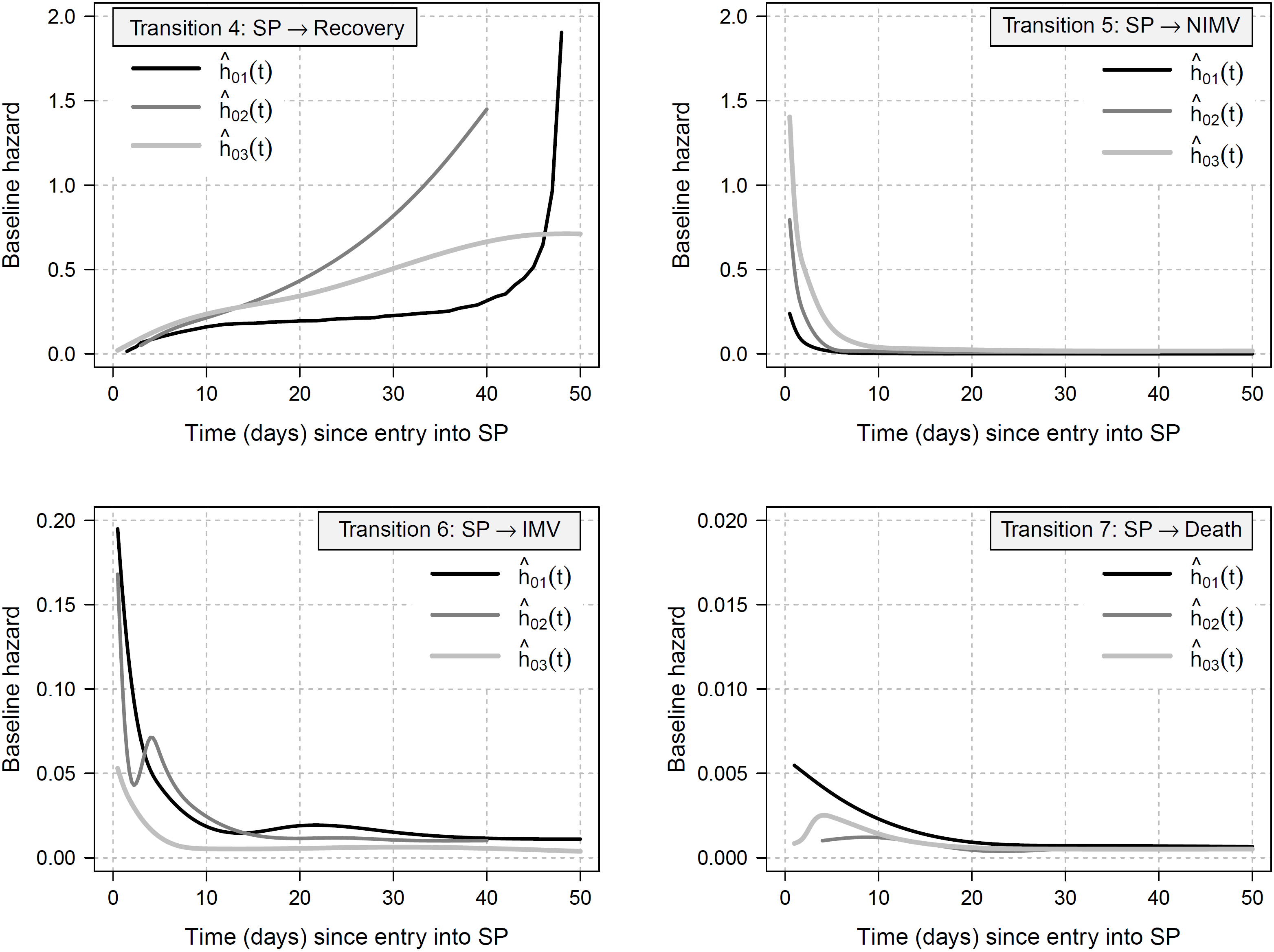}
\vspace{-0.35cm}
\end{center}
\caption{Smoothed estimates for the transition-specific baseline 
hazards per cohort using the M2~approach in the \textsf{div3W}~data. 
Baseline hazards are particularly estimated for all subjects moving 
from the SP state towards subsequent states, regardless the time 
point $t_\tx{SP}$ at which a subject has entered into SP.
\label{F4}}
\end{figure}
\subsection{Forecasting time-to-event outcomes for subject-specific profiles}
Until now, the results have been largely discussed in terms of both covariate 
and cohort effects on a given transition hazard, based on either the M1 and 
M2 multistate modeling approaches. While these results are quite interesting 
by themselves, the multistate modeling framework also allows us to answer 
clinically important subject-specific questions involving multiple transitions, 
while conditioning on a set of personal characteristics. In particular, one 
of the most attractive features of the general multistate modeling framework 
relies on using the hazard transition estimates to compute individualized 
predictions for the event history process. These predictions for a subject's 
prognosis are indeed expected to be highly accurate since intermediate states 
in the whole process can be accounted for. Let us consider a fictitious 
subject, say $k$, from our target population but not included in our motivating 
dataset, known to occupy the $\ell$ state at time $s \geq 0$. We are interested 
in computing the subject-specific conditional probability of occupying a future 
state, say $m$, given a set of prognostic features at hospital entry, $\bz_k$, 
the time of entry into the current state, $t_{\ell k}$, and the \mbox{COVID-19} 
wave, $g = 1, 2, 3$:
\begin{equation*}
P_k^\lm (s, t) = \Pr \{ X_k(t) = m \mid X_k(s) = \ell, \E 
  \bz_k, \E t_{\ell k}, \E c_{gk} = 1 \},~~ s < t. \\[1.0pt]  
\end{equation*}
When the Markov assumption does not hold in all transitions, 
empirical inferences regarding prediction probabilities can 
be still consistently derived from the non-parametric AJ 
estimator, given that censoring is not related to the states 
occupancy or to the transition times between states (Datta 
\& Satten, 2001). Thus, the procedures implemented in the 
\textsf{R} package \cx{mstate} remain valid for our purposes. 
Alternatively, using the same package, prediction probabilities 
can be performed by simulating a sufficiently large number of 
trajectories, say $M = 10000$, through the corresponding 
multistate setting.

For illustration, we retrospectively explore the evolution 
of two fictitious subject profiles from our target population. 
Particularly, let us consider two distinct medical profiles 
for a 50-year-old male, one low-risk and the other high-risk, 
whose health forecasting is expected to be vastly different. 
Regarding the safi covariate, the uncentered value for the 
low-risk profile is 476.2 units, whereas the high-risk profile 
has 410 units. Both subject trajectories are assumed to start 
from the SP state at time zero, while their quantitative 
covariates are accounted for after being centered around their 
corresponding means. We then focus on understanding how each 
profile's medical information affects the probability of 
occupying any subsequent state during the hospitalization period. 
Of special relevance is the mortality forecasting of a subject, 
which in our multistate scheme is equivalent to estimating the 
probability of dying in the hospital.
\begin{figure}[!ht]
\begin{center}
\includegraphics[scale = 0.150]{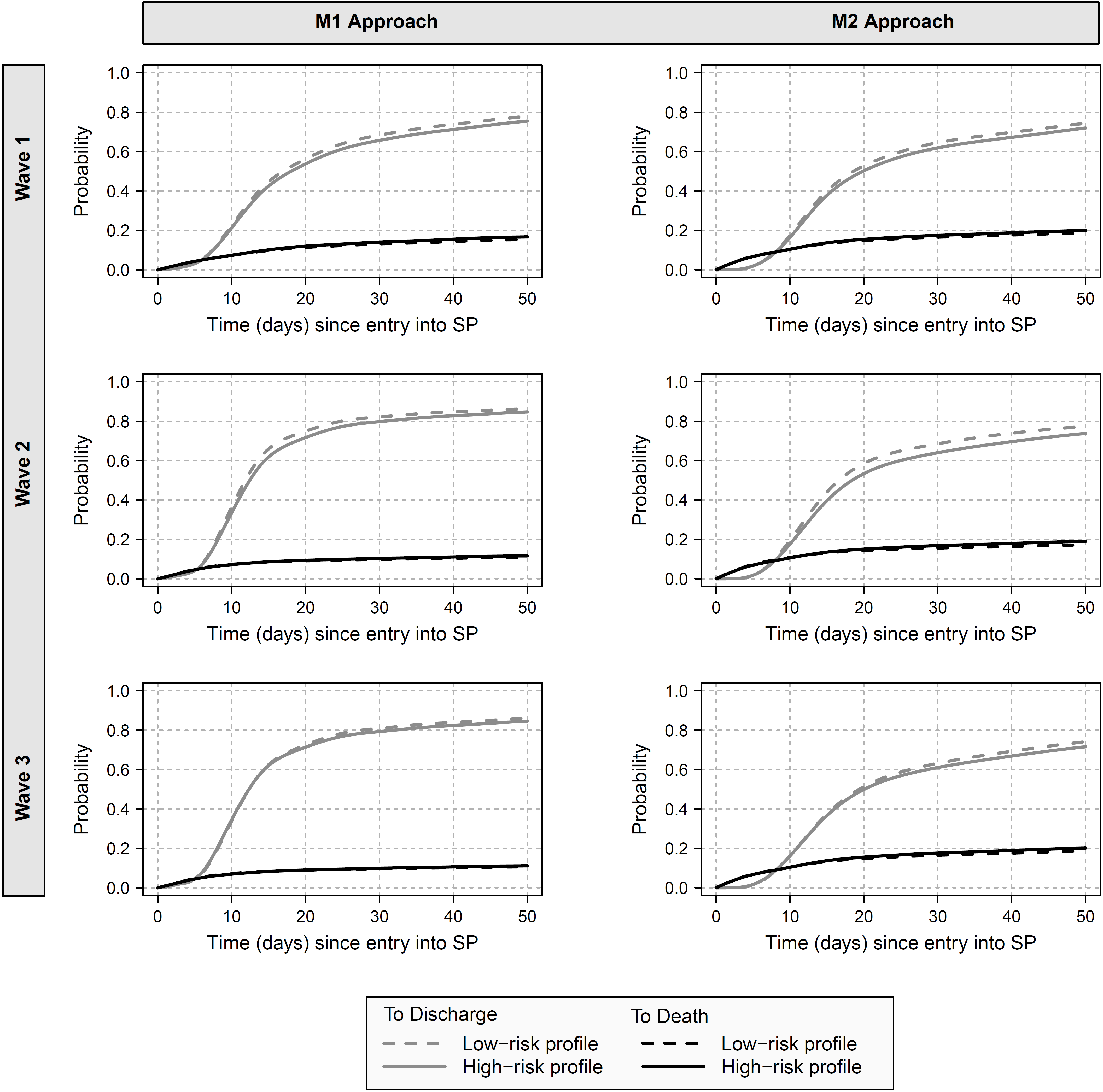}
\end{center}
\caption{Smoothed estimates for the prediction probabilities 
related to the low-risk and high-risk profiles considered. Results 
are derived from both the M1 and M2 approaches and each of the three 
\mbox{COVID-19} waves.\label{F5}}
\end{figure}
\par
For the fictitious low-risk and high-risk profiles, Figure~5 
shows the prediction probabilities computed from the M1 and M2 
approaches within each of the three \mbox{COVID-19} waves. In 
essence, the subject's baseline features seem to have an 
influence when forecasting mortality risk, as evidenced in 
all the possible combinations of both the approach employed 
and the wave of reference. For a given time $t$ at which a 
prediction is to be made, a low-risk subject profile 
systematically gives rise to a higher probability of being 
discharged than the high-risk profile, and this relationship 
between profiles is just the opposite in regard to the 
probability of dying in the hospital. Equally important is 
quantifying the effect of the wave covariate when estimating 
prediction probabilities, in which the first wave yields smaller 
differences between survivors and nonsurvivors within the M1 
approach. A plausible explanation for this pattern must be 
found in the limited knowledge of the disease in the first 
wave, giving rise to a higher number of in-hospital deaths in 
comparison with second- and third-wave pandemic periods. Also 
note that, for any of the waves under M2, differences between 
survivors and nonsurvivors clearly diminish with respect to M1. 
Therefore, it is clear the influence of the wave covariate 
itself to predict state occupation probabilities.

To quantify the departure between the probability of being 
discharged and dying, the prediction probabilities for low- and 
high-risk profiles can be evaluated at distinct time points after 
experiencing SP. In our data, subjects entering the hospital at 
SP show a median overall stay of 15 days (mean of 22.5 days), so 
$t = 10$ and $t = 20$ days are reasonable choices. Prediction 
results can be consulted in Table~S7 of Appendix~C.

To conclude this section, another point worth mentioning is the dynamic 
behavior of the prediction probabilities. In practice, this means that 
future trends for a particular subject profile can be updated as new 
information becomes available during the hospitalization period. To 
exemplify this dynamic forecasting using the two risk profiles considered, 
Figures~S23$\E$--$\E$S24 of Appendix~C shows the evolution of the prediction 
probabilities as a given subject profile enters later in the SP state. 
Hence, given a particular low-risk and high-risk subject profile who entered 
into the SP state at $t \in \{0, 5, 10\}$ days, the probabilities of either 
being discharged or dying are computed for the M1 and M2 approaches.
\section{Discussion}
A covariate-cohort approach, M1, and a stratum-cohort approach, M2, 
have been proposed as Cox-based multistate modeling strategies to 
analyze event history data from different cohorts. These approaches 
postulate alternative ways to properly accommodate the 
heterogeneity across distinct subpopulations, since subjects 
belonging to the same cohort tend to share unobserved characteristics 
that cannot be explicitly accounted for. The M1 covariate-cohort 
approach considers a collection of cohort-specific binary indicators, 
serving as proxies for collecting within-cohort unobserved characteristics. 
As an alternative strategy, the M2 stratum-cohort approach delves deeper 
into explaining within-cohort subject behavior by means of allowing 
an extra flexibility to estimate transition-specific underlying risks 
for different cohorts. While M1 is more often preferred when the 
cohort-specific covariate effects are of main interest, M2 is more 
likely to be chosen when the research focus is on transition probabilities. 
A further advantageous aspect to mention in M2 approach is that 
stratifying the event history data by group eases compliance with the 
proportional hazards hypothesis. Nonetheless, the cohort effect can be 
neither estimated nor statistically tested, since it is used for 
stratification. On another note, the Markov property conditional on 
covariates has been assessed using a global score test. Whenever a 
non-Markov environment is detected, the time of entry into the current 
state are included in the corresponding Cox model, setting our analysis 
into a semi-Markov framework. The relevance of this time-to-entry 
covariate is easily evaluated by performing a likelihood ratio test 
to assess whether the corresponding regression parameter is different 
from zero. Based on this rationale, the two testing procedures applied 
should be understood as complementary: the global score test serves as 
a pre-test for the Markovian condition, while the incorporation of the 
time-of-entry covariate offers a simple alternative for effectively 
considering a specific aspect of previous history. 

The two suggested approaches have been applied to analyze a large 
sample of \mbox{COVID-19-hospitalized} subjects grouped into three 
distinct pandemic waves, all occurring in the southern Barcelona 
metropolitan area during the prevaccine era. The assessment of the 
Markov property in our multistate process revealed non-Markovianity 
for the transitions from severe pneumonia to noninvasive mechanical 
ventilation and invasive mechanical ventilation. As a consequence, 
the time of entry into the SP state,~\tsp, has been incorporated as 
an explanatory covariate when modeling the transition hazard. The 
relationship between~\tsp~and the transition hazard led us to 
conclude that later entry times into the SP state or, equivalently, 
shorter sojourn times, are strongly associated with a higher risk of 
requiring mechanical ventilation, either NIMV or IMV. In parallel, 
the presumed dependence among subjects in the same wave has been 
inspected using the two previously mentioned strategies for accommodating 
wave influences. Concerning the impact of the covariates on a given 
transition, parameter estimates under the M1 fixed-effects approach 
appear to parallel those derived under the M2 wave-stratified approach, 
and the same can be said for the corresponding standard errors. In 
fact, in the near absence of non-proportional wave-specific baseline 
hazards, these practically equivalent estimates could be anticipated 
since both approaches assume a multiplicative relationship between 
the covariates and the hazard function.

The analysis of our empirical \mbox{COVID-19}~data involves distinct 
critical considerations that must be pointed out. First, our three 
waves were not randomly sampled, but were rather gathered during a 
very specific historical disease period, within a particular 
geographic area. Hence, for more general conclusions to be drawn, 
it would be interesting to consider data in other geographical areas 
during the same disease period. Second, the different disease patterns 
observed across waves are largely due to the particular hospital 
protocol adopted within a given pandemic period (Straw et al., 
2021).~More specific guidance would therefore be necessary in this 
respect to better understand the ceiling of care criterion 
employed in each period. For instance, one additional source of 
information could be hospital-bed occupancy rates~(Bekker et 
al., 2023).~Additionally, the treatments employed have also 
changed over the course of the pandemic; a paradigmatic 
example is the use of corticosteroids, whose extended administration 
to hospitalized subjects since the second wave led to a large 
improvement in the effectiveness of \mbox{COVID-19} treatment 
(Balaz et al., 2021).~In our case, a careful comparison between 
the characteristics of wave-specific subjects has enabled us to 
explain, whether through the M1 or M2 approach, the later 
differences observed in the behavior of each wave. A further 
key point of our dataset is the time that a subject recovers 
from \mbox{COVID-19}, which is actually unobserved. Instead, 
it has been assumed that recovery occurs two days prior to 
discharge. An alternative would consist of considering a 
latent recovery state, which would be omitted when fitting 
the Cox-based multistate models. In practice, this would 
entail the disappearance of the recovery~$\to$~discharge 
and recovery~$\to$~death transitions, leaving the same number 
of subjects moving in transitions that did not involve the 
recovery state. Among the transitions with the required sample 
size for obtaining parameter estimates, only in the NSP~$\to$~SP 
and NSP~$\to$~discharge transitions the results would remain 
unchanged.

In addition to explaining subject-specific disease trajectories, we 
have obtained individualized prediction probabilities from the model 
fitted under both approaches. The presence of semi-Markov effects 
obviously makes this more difficult, although the non-parametric AJ 
estimator still guarantees consistent estimates. Thus, although 
the two approaches cannot be directly compared, it is possible to 
graphically compare the estimates for the state occupation 
probabilities. As an interesting alternative,~Crowther and Lambert 
(2017)~use a simulation-based method in the context of parametric 
models, wherein a wide variety of candidate models is available. A 
detailed comparison of the inference performance of both procedures 
could be a valuable line of future research to reach a better 
understanding of subject trajectories over the entire multistate process.

In conclusion, the identified trade-offs between the M1 and M2 
approaches do not necessarily imply raising the question of 
whether one of the approaches is preferable to the other. On 
the contrary, we could argue that this is not a dichotomous 
choice at all, since both semi-Markovian approaches represent 
reliable and easy-to-implement alternatives to analyze 
multicohort data in the multistate framework. More generally, 
given a certain dataset to be analyzed, the selection of one 
or the other approach will depend on both which cohort-specific 
key aspects are to be highlighted and the focus of the inference. 


\section*{Acknowledgments}
This research was partially supported by grants from the Spanish Ministry of 
Science and Innovation (PID2019-104830RB-I00) and from the Catalan Department 
of Research and Universities (2020PANDE00148). The authors are very grateful 
to Professor A. C. Titman for his outstanding help in clarifying some aspects 
of the Cox-based global test for checking the Markov hypothesis. The members 
of the Divine project to identify COVID-19 risk factors are the following 
(alphabetically ordered by last name): G. Abelenda-Alonso, M. Besalú, 
\E J. Carratalà, \E E. Cobo, \E J. Cortés, \E D. Fernández, \E L. Garmendia, 
\E G. Gómez Melis, \E C. Gudiol, \\ P. Hereu, K. Langohr, G. Molist, N. Pallarès, 
N. Pérez-Álvarez, X. Piulachs, A. Rombaut, C. Tebé, and S. Videla.

\section*{Conflict of interest}
\vspace{-0.25cm}
The authors declare no potential conflicts of interest.

\section*{Supporting information}
\vspace{-0.25cm}
The web-based supplementary information contains the tables and figures 
reported by the aforementioned appendices A, B, and C. Moreover, the 
\textsf{R} code implementing our two modeling approaches are provided 
in Appendix~D. Particularly, Section~D.1 provides the code to perform 
the global testing procedure for the Markov property, Section~D.2 the 
code for parameter estimates, Section~D.2 the code for hazard ratio 
estimates, and Section~D.4 shows how to compute predictions.


\nocite{*}

\begin{thebibliography}{10}

\bibitem{andersen2002}
Andersen,~A., \& Keiding,~N.~(2002).~Multi-state models for 
event history analysis.~{\it Statistical Methods in Medical 
Research, 11}(2)\string, 91--115.


\bibitem{houwelingen2008}
van Houwelingen,~W.C., \& Putter,~H.~(2008).~Dynamic 
predicting by landmarking as an alternative for multi-state 
modeling: An application to acute lymphoid leukemia data.
~{\it Lifetime Data Analysis, 14}(4)\string, 447--463.


\bibitem{meira2009}
Meira-Machado,~L., de Uña-Álvarez,~J., Cadarso-Suárez,~C., 
\& Andersen,~P.K.~(2009).~Multistate models for the analysis 
of time-to-event data.~{\it Statistical Methods in Medical 
Research, 18}(2)\string, 195--222.


\bibitem{geskus2015}
Geskus, R.B.~(2015).~{\it Data Analysis with Competing Risks 
and Intermediate States}.\newblock Chapman \& Hall/CRC: Boca 
Raton, Florida (USA). 


\bibitem{cook2018}
Cook, R.J. \& Lawless, J.F.~(2018).~{\it Multistate Models 
for the Analysis of Life History Data}. \newblock Chapman 
\& Hall/CRC: Boca Raton, Florida (USA).


\bibitem{darroch1990}
Darroch,~J.N. \& McCloud,~P.I.~(1990).~Separating two 
sources of dependence in repeated influenza outbreaks.~{\it 
Biometrika, 77}(2)\string, 237--243.


\bibitem{birkenbihl2022}
Birkenbihl,~R.J., Salimi,~S., Frohlich,~H., Japanese 
Alzheimer's Disease Neuroimaging Initiative, \& Alzheimer's 
Disease Neuroimaging Initiative.~(2022).~Unraveling the 
heterogeneity in Alzheimer's disease progression across 
multiple cohorts and the implications for data-driven 
disease modeling.~{\it Alzheimer's \& Dementia, 18}(2)
\string, 251--261.


\bibitem{freijser2023}
Freijser,~L., Annear,~P., Tenneti,~N., Gilbert,~K., 
Chukwujekwu,~O., Hazarika,~I., \& Mahal,~A.~(2023).~The 
role of hospitals in strengthening primary health care 
in the Western Pacific.~{\it The Lancet Regional 
Health - Western Pacific, 33}\string, 1--8. 


\bibitem{gelman2007}
Gelman,~A., \& Hill,~J.~(2007).~{\it Data Analysis Using 
Regression and Multilevel/Hierarchical Models}. \newblock 
Cambridge University Press, Cambridge (UK).


\bibitem{huang2020}
Huang,~C, Wang,~Y, Li,~X, Ren,~L., Zhao,~J., Hu,~Y., \ldots 
Cao,~B.~(2020).~Clinical features of patients infected with 
2019 novel coronavirus in Wuhan,China.~{\it The Lancet, 
395}(10223)\string, 497--506.


\bibitem{chan2020}
Chan,~J.F., Yuan,~S., Kok~K.-H., Kai-Wang To,~K., Chu,~H., 
Yang,~J., \ldots, Yuen,~K.-Y.~(2020).~A familial cluster 
of pneumonia associated with the 2019 novel coronavirus 
indicating person-to-person transmission: a study of a 
family cluster.~{\it The Lancet, 395}(10223)\string, 
514--523.


\bibitem{carbonell2021}
Carbonell,~R., Urgeles,~S., Rodríguez,~A., Bodí,~M., 
Martín-Loeches,~I., Solé-Violán,~J., \ldots, COVID-19 
SEMICYUC Working Group.~(2021).~Mortality comparison 
between the first and second/third waves among 3,795 
critical \mbox{COVID-19} patients with pneumonia 
admitted to the ICU: A multicentre retrospective 
cohort study.~{\it The Lancet Regional Health - Europe, 
11}(100243)\string, 1--9.


\bibitem{buttenschon2022}
Buttenschon,~H.N., Lynggaard,~V., Sandbøl,~S.G., 
Glassou,~E.N., \& Haagerup,~A.~(2022).~Comparison of 
the clinical presentation across two waves of 
\mbox{COVID-19}: A retrospective cohort study.~{\it BMC 
Infectious Diseases, 22}(423)\string, 1--11.


\bibitem{pallares2023}
Pallarès,~N., Tebé,~C., Abelenda-Alonso,~G., Rombauts,~R., 
Oriol,~I., Simonetti,~A.F., \ldots, MetroSud and Divine 
study groups.~(2023).~Characteristics and outcomes by 
ceiling of care of subjects hospitalized with COVID-19 
during four waves of the pandemic in a metropolitan area: 
a multi-center cohort study.~{\it Infectious Diseases and 
Therapy, 12}(1)\string, 273--289.


\bibitem{wendel2021}
Wendel-Garcia,~P.D., Aguirre-Bermeo,~H., Buehler,~P.K., 
Alfaro-Farias,~M., Yuen,~B., David,~S., \ldots, 
Ristic,~A.~(2021).~Implications of early respiratory support 
strategies on disease progression in critical \mbox{COVID-$19$}: 
a matched subanalysis of the prospective RISC-19-ICU cohort.~{\it 
Critical Care, 25}(1)\string, 1--12.


\bibitem{berenguer2021}
Berenguer,~J., Borobia,~A.M., Ryan,~P., Rodríguez-Baño, J., 
Bellón,~JM, Jarrín,~I., \ldots, COVID@HULP Working Group.~(2021).~
Development and validation of a prediction model for 30-day 
mortality in hospitalised patients with \mbox{COVID-19}: the 
\mbox{COVID-19} SEIMC score.~{\it Thorax, 76}(9)\string, 920--929.





 
\bibitem{catoire2021}
Catoire,~P., Tellier,~E., de la Rivière,~C., Beauvieux,~M.C., 
Valdenaire,~G., Galinski,~M., Revel,~P., Combes,~X., \& 
Gil-Jardiné,~C.~(2021).~Assessment of the SpO2/FiO2 ratio 
as a tool for hypoxemia screening in the emergency department.
~{\it The American Journal of Emergency Medicine, 44}(23)
\string, 116--120.


\bibitem{aalen1978}
Aalen,~O.O., \& Johansen,~S.~(1978).~An Empirical Transition 
Matrix for Non-Homogeneous Markov Chains Based on Censored 
Observations.~{\it Scandinavian Journal of Statistics, 5}(3)
\string, 141--150.


\bibitem{cox1972}
Cox,~D.R.~(1972).~Regression models and life tables (with 
Discussion).~{\it Journal of the Royal Statistical Society: 
Series B, 34}(2)\string, 187--220.


\bibitem{andersen1991}
Andersen,~P.K., Hansen.,~L.S., \& Keiding,~N.~(1991).~Non- 
and semi-parametric estimation of transition probabilities 
from censored observation of a non-homogeneous Markov 
process.~{\it Scandinavian Journal of Statistics, 18}(2)
\string, 153--167.


\bibitem{meira2006}
Meira-Machado,~L., de U\~na-Álvarez,~J., \& 
Cadarso-Suárez,~C.~(2006).~Nonparametric estimation of 
transition probabilities in a non-Markov illness-death model.
~{\it Lifetime Data Analysis, 12}(3)\string, 325--344.


\bibitem{andersen2008}
Andersen,~A., \& Perme,~M.P.~(2008).~Inference for outcome 
probabilities in multi-state models.~{\it Lifetime Data 
Analysis, 14}(4)\string, 405--431.


\bibitem{titman2015}
Titman,~A.~(2015).~Transition Probability Estimates for 
Non-Markov Multi-State Models.~{\it Biometrics, 71}(4)
\string, 1034--1041.


\bibitem{breslow1972}
Breslow,~N.E.~(1972).~Contribution to the Discussion 
of the paper by D. R. Cox.~{\it Journal of the Royal 
Statistical Society: Series B, 34}(2)\string, 216--217.


\bibitem{putter2007}
Putter,~H., Fiocco,~M., \& Geskus,~R.B.~(2007).~Tutorial 
in biostatistics: Competing risks and multi-state models. 
~{\it Statistics in Medicine, 26}(11)\string, 2389--2430.


\bibitem{ieva2017}
Ieva,~F., Jackson,~C.H., \& Sharples,~L.D.~(2017).~Multi-state 
modelling of repeated hospitalisation and death in patients 
with heart failure: The use of large administrative databases 
in clinical epidemiology.~{\it Statistical Methods in Medical 
Research, 26}(3)\string, 1350--1372.


\bibitem{wreede2010}
de Wreede,~L.C., Fiocco,~M., \& Putter,~H.~(2010). The 
\textsf{mstate} Package for Estimation and Prediction in 
Non- and Semi-Parametric Multi-State and Competing Risks 
Models.~{\it Computer Methods and Programs in Biomedicine, 
99}(3)\string, 261--274.


\bibitem{una2015}
de U\~na-Álvarez,~J., \& Meira-Machado,~L.~(2015). 
Nonparametric Estimation of Transition Probabilities 
in the Non-Markov Illness-Death Model: A Comparative 
Study.~{\it Biometrics, 71}(2)\string, 364--375.


\bibitem{putter2018}
Putter,~H., \& Spitoni,~C.~(2018).~Non-parametric estimation 
of transition probabilities in non-Markov multi-state models: 
The landmark Aalen–Johansen estimator.~{\it Statistical 
Methods in Medical Research, 27}(7)\string, 2081--2092.


\bibitem{titman2022}
Titman,~A.C., \& Putter,~H.~(2022).~General tests of the 
Markov property in multi-state models.~{\it Biostatistics, 
23}(2)\string, 380--396.


\bibitem{mammen1993}
Mammen,~E.~(1993).~Bootstrap and Wild Bootstrap for 
High Dimensional Linear Models.~{\it Annals of Statistics, 
21}(1)\string, 255--285.


\bibitem{vittinghoff2007}
Vittinghoff,~E., \& McCulloch,~C.E.~(2007).~Relaxing 
the Rule of Ten Events per Variable in Logistic and Cox 
Regression.~{\it American Journal of Epidemiology, 165}(6)
\string, 710--718.


\bibitem{lin1993}
Lin,~D.Y., Wei, L.J., \& Ying, Z.~(1993).~Checking the Cox model 
with cumulative sums of martingale-based residuals.~{\it Biometrika, 
80}(3)\string, 557--572.


\bibitem{martinussen2006}
Martinussen,~T., \& Scheike, T.H.~(2006).~{\it Dynamic Regression
Models for Survival Data. Statistics for Biology and Health.}\newblock 
Springer, New York (USA). 


\bibitem{datta2001}
Datta,~S., \& Satten,~G,A.~(2001).~Validity of the 
Aalen–Johansen estimators of stage occupation probabilities 
and Nelson–Aalen estimators of integrated transition hazards 
for non-Markov models.~{\it Statistics \& Probability Letters, 
55}(4)\string, 403--411.

\bibitem{straw2021}
Straw,~S., McGinlay,~M., Drozd,~M., Slater, T.A., Cowley, A., 
Kamalathasan, S., \ldots, Witte, K.K.~(2021).~Advanced care 
planning during the \mbox{COVID-19} pandemic: ceiling of 
care decisions and their implications for observational 
data.~{\it BMC Palliative Care, 20}(10)\string, 1--11.


\bibitem{bekker2023}
Bekker~R., uit het Broek,~M., \& Koole, G.~(2023).~Modeling 
COVID-19 hospital admissions and occupancy in the Netherlands.
~{\it European Journal of Operational Research, 304}(1)
\string, 207--218.


\bibitem{balaz2021}
Balaz,~D, Wikman-Jorgensen,~PE, Giner,~V, Rubio-Rivas,~M.,
de Miguel,~B., Noureddine,~M., \ldots, Casas-Rojo,~J.M.~
(2021).~Evolution of the Use of Corticosteroids for the 
Treatment of Hospitalised \mbox{COVID-19} Patients in Spain 
between March and November 2020: SEMI-COVID National 
Registry.~{\it Journal of Clinical Medicine, 10}(19)
\string, 1--17.


\bibitem{crowther2017}
Crowther,~M.J., \& Lambert,~P.C.~(2017).~Parametric 
multistate survival models: Flexible modelling allowing 
transition-specific distributions with application to 
estimating clinically useful measures of effect 
differences.~{\it Statistics in Medicine, 36}(29)
\string, 4719--4742.

\end{thebibliography}
\end{document}